\documentclass[preprint2,times,tighten]{aastex6}

\usepackage{color}
\usepackage{enumitem}
\usepackage{hyperref}
\usepackage{verbatim}
\usepackage{amsmath,amstext,mathrsfs}
\usepackage[all]{hypcap} 
\usepackage{afterpage}
\usepackage{color}

\newcommand*{\kh}{} 

\newcommand*{\vl}{}

\begin{document}
\title{Gradients of Synchrotron Polarization: Tracing 3D distribution of magnetic fields}

\author{A. Lazarian\altaffilmark{1}, Ka Ho Yuen\altaffilmark{1}}
\email{lazarian@astro.wisc.edu }
\altaffiltext{1}{Department of Astronomy, University of Wisconsin-Madison}

\begin{abstract}
We describe a new technique for probing galactic and extragalactic 2D and 3D magnetic field distribution using gradients of polarized synchrotron emission. The fluctuations of magnetic field are elongated along the ambient magnetic field. Therefore, the field variations are maximal perpendicular to the B-field. This allows tracing B-field with synchrotron polarization gradients (SPGs). We demonstrate that the Faraday depolarization allows to map 3D B-field structure by. The depolarization ensures that the polarization gradients sample the regions close to the observer with the sampling depth controlled by the frequency of radiation.  We also analyze the B-field properties along the line-of-sight by applying the gradient technique to the wavelength derivative of synchrotron polarization. This Synchrotron Derivative Polarization Gradients (SDPGs) technique can recover the 3D vectors of the underlying B-fields. The new techniques are different from the Faraday tomography as they provide a way to map the 3D distribution of B-fields components perpendicular to the line of sight. In addition, we find that the alignment of gradients of polarization with the synchrotron polarization can be used to separate the contribution of the foreground from the polarization of cosmological origin. We notice that the same alignment is also present for the dust polarization.
\end{abstract}
\section{Introduction}

Interstellar media (ISM) of spiral galaxies are known to be both magnetized and turbulent (see  \citealt{Armstrong1995ElectronMedium, Chepurnov2010ExtendingData}), with turbulent magnetic fields playing a critical role in many key astrophysical processes. These include star formation (see \citealt{MO07,MK04}), the propagation and the acceleration of cosmic rays (see \citealt{J66,YL08})  and the regulation of heat and mass transfer between different ISM phases (see \citealt{D09} for the list of the different ISM phases). In addition, the study of enigmatic CMB B-modes \citep{1997PhRvD..55.1830Z} is impeded by the interference of fluctuations arising from galactic magnetic field 
\citep{2017PhRvL.118i1801C,2017MNRAS.464.3617K,2017MNRAS.470.3103K}. Thus the detailed knowledge of the interstellar magnetic field structure is essential for both astrophysical and cosmological studies. 
 
However, measuring the properties of magnetic field is notoriously difficult. There are only a few techniques for studying the interstellar magnetic fields that {\toreferee is successful} for probing magnetic fields in low-latitude diffuse ISM and molecular clouds. Zeeman splitting directly measures the strength of the line-of-sight (LOS) component of magnetic fields in molecular clouds \citep{2012ARA&A..50...29C,2015aska.confE.110R}. Despite the long integration time employed to observe its weak effect, Zeeman measurement can only access the high-magnitude end of the interstellar magnetic fields in a mildly turbulent environment and can provide only limited morphology information. The morphology of the plane-of-sky component of magnetic fields can be obtained through starlight polarization (e.g. \citealt{2011ApJ...740...21P}) and polarized dust grain emissions (e.g. Planck collaboration 2015).  Besides the limitations or even failures arising from poorly controlled variations in radiation anisotropies, uncertain dust grain properties, {\toreferee as} well as the effects of large optical depths (see \citealt{2015ARA&A..53..501A}), the polarization observations are in general difficult because the polarization percentage from telescope observations is comparable to the polarization arising from interstellar dust \citep{2015ASPC..499..197G}. The detection of magnetic fields in the high-latitude diffuse ISM is even more difficult, as it requires a higher sensitivity that is beyond the ability of present day instruments. 

Measurements of polarized synchrotron radiation and Faraday rotation (see \citealt{2011MNRAS.412.2396F,2013pss5.book..641B,2015A&A...575A.118O,2016ApJ...830...38L,2017A&A...597A..98V}) provide an important insight into the magnetic structure of the Milky Way and neighboring galaxies. More synchrotron data is being available with new instruments. For instance, Low Frequency Array (LOFAR) (see \citealt{vH13}) provides insight into low frequency synchrotron for which Faraday depolarization may be important. The value of synchrotron as the source of information is only going to increase as the Square Kilometer Array will provide detailed maps of diffuse emission with unprecedented resolution (see \citealt{2015ApJ...811...40S,2015aska.confE..92J}). This motivates us to study new ways how the synchrotron data can be used. 

In what follows we deal with synchrotron polarization gradients \citep{Gaensler2011Low-Mach-numberGradients,Burkhart2012PropertiesMaps,2014A&A...566A...5I,2014MNRAS.437.2936S,2015MNRAS.451..372R,2017MNRAS.466.2272H}, but in this paper we use them in a way different from the earlier studies. This work continues our exploration of gradients {of synchrotron polarization\toreferee under the framework of Velocity Gradient Technique (VGT)} as a means of tracing magnetic fields in different media. {\toreferee VGT was} introduced in a series of recent papers \citep{GL17, YL17a,YL17b,LY18a}, {\toreferee which} is based on the modern understanding of MHD turbulence cascade (\citealt{GS95}, henceforth GS95, see also \citealt{2013SSRv..178..163B} for a review) and turbulent reconnection (\citealt{Lazarian1999ReconnectionField}, henceforth LV99, \citealt{2011ApJ...743...51E}, see \citealt{2015ASSL..407..311L} for a review). In particular, it is important that velocity eddies are aligned with the surrounding {\it local}\footnote{This property can be understood easily by considering the response of magnetized fluid by considering the force balancing between the randomness of turbulence motions compared to stiffness of magnetic field lines. If magnetic field is dynamically important, it is more difficult to bend magnetic field lines than to mix them perpendicularly with respect to their directions. These mixing motions are not constrained as magnetic reconnection in turbulent fluids is fast, as predicted by the theory of turbulent reconnection (\citealt{Lazarian1999ReconnectionField}). As a result, the eddies perpendicular to the magnetic field in their vicinity have hydrodynamic-like nature and the Kolmogorov spectrum. In other words, the gradients of the velocities are perpendicular to the magnetic field. }  magnetic field (LV99, \citep{CV00,2001ApJ...554.1175M}) The VGT has proven to be a new promising way of studying magnetic field in both galactic HI and in molecular clouds (see \citealt{YL17b,LY18a})

For Alfvenic turbulence, magnetic fluctuations and velocity fluctuations are symmetric. This opens ways for magnetic field tracing {\toreferee by} studying {\toreferee the} gradients of {\toreferee observables of }magnetic fields {\toreferee, e.g. Synchrotron radiations. Synchrotron intensities and polarized intensities resulted by Faraday rotation} are the primary informants about magnetic field fluctuations. 

The closest to the new technique that we introduce in this paper is the one in \citeauthor{LYLC17} (\citeyear{LYLC17}, LYLC) where we proposed magnetic field tracing with {\it synchrotron intensities}. Unlike the traditional technique of using synchrotron polarization, the new technique employed the {\it synchrotron intensity gradients} (SIGs) to trace magnetic field in interstellar media. The advantage of the SIGs compared to the tracing of magnetic field using {\toreferee orientations of }synchrotron polarization is that they do not require polarization studies and, more importantly, do not require the correction of the Faraday rotation effect. This technique was applied in LYLC to the Planck synchrotron intensity maps and the results were shown to be in good agreement with the Planck magnetic field maps obtained using synchrotron polarization. 
 
The present paper studies the ways of getting 3D distribution of magnetic fields {\it {\toreferee by} synchrotron polarization gradients}. In previous works, we described how to use the galactic rotation curve \citep{GL18} and multi-molecular tracers (Hu et. al, in prep) to map magnetic field in 3D (See \S \ref{sec:dis}). The present paper explores the depolarization arising from the Faraday rotation to map 3D interstellar magnetic field structure.  Our study is focus on describing 2 new techniques, namely, the Synchrotron Polarization Gradients (SPGs, \S \ref{sec:limiting} \& \S \ref{sec:spg-meanb}  ) technique and the one that employs gradients of the wavelength derivative of synchrotron polarization, i.e. Synchrotron Polarization derivative Gradients (SPDGs, \S \ref{sec:FRG}).   We note that the aforementioned techniques are different from the Faraday tomography (\citealt{1966MNRAS.133...67B,BB05}, {\torefereetwo see  \citealt{2013A&A...558A..72I,2014A&A...568A.101J,2015A&A...583A.137J,2017A&A...597A..98V}} for recent observation in LOFAR polarimetric observations) in a few ways. First of all, SPGs are capable to study the magnetic field structure perpendicular to the line of sight. In addition, the both SPGs and SPDGs map the distribution of magnetic fields is obtained in real space rather than in terms of the Faraday polarization depths. It is also important that, compared to the Faraday Tomography, our techniques do not require the measurements at hundreds of frequencies\footnote{\toreferee The tracing power of SPG is proportional to the number of available slices in the position-position-frequency map $N$ instead of $\propto \sqrt{N}$ in Faraday tomography. For instance, for the same 100 frequencies SPG provides 100x better sensitivity. The synergy is coming from the fact that the techniques measure different components of magnetic field, the SPGs gets the POS component of B-field and the Faraday tomography deals with the LOS magnetic field component.} in order to acquire a relatively accurate result. 

For our treatment of the gradients the guidance is provided the theory of polarized synchrotron fluctuations formulated in \citeauthor{LP16} (\citeyear{LP16}, hereafter LP16) A number of predictions of LP16 were tested in the earlier papers (see \citealt{2016ApJ...831...77L,2016ApJ...825..154Z}). In our present paper, a few more LP16 predictions are tested, i.e. see (\S \ref{sec:limiting} for the structure dependence of correlation {\vl and} \S \ref{sec:LP18} for the power law prediction of LP16). 

In \S \ref{sec:principle}{\vl ,} we describe the fundamental principle of the gradient technique. In \S \ref{sec:theory}, we briefly describe the theoretical foundation of SPGs. In \S \ref{sec:numerics}, we discuss the MHD simulations that we use to test our predictions. In \S \ref{sec:limiting}, we discuss the behavior of SPGs in the case of random-field Faraday rotation. In \S \ref{sec:FRG}{\vl ,} we discuss the estimation of Faraday Rotation measure and the relation to the synchrotron polarization derivative in respect to squared wavelength. In \S \ref{sec:spg-meanb} we discuss how Faraday Rotation effect due to a mean field along the line of sight would change the properties of SPGs. In \S \ref{sec:recipe3d},  we establish the recipe of constructing 3D magnetic field from the gradient technique. We provide the discussion of our results in \S \ref{sec:dis} and {\vl summarize} in \S \ref{sec:con}. As for the Appendix, in \S \ref{sec:LP18} we show the statistical testing {\vl of} LP16, in \S \ref{sec:Zgrad} we illustrate {\vl that} the properties of the Zeeman measure gradients are similar to those of the Faraday rotation measure. In \S \ref{dust_pol} we discuss about the possibilities of using gradients of dust polarizations as we did in synchrotron polarizations, and in \S \ref{sec:spdg} we discuss our preliminarily recipe on the 3D magnetic field based on SPDGs.

\section{The fundamentals of gradient technique}
\label{sec:principle}

As we mentioned earlier, physically one can imagine the strong MHD turbulence (GS95) as a superposition of anisotropic eddies aligned with magnetic field. Since turbulent reconnection (LV99) enables unconstrained mixing motions in the direction perpendicular to magnetic field, the gradients of eddy structures are thus perpendicular to the local magnetic field directions. As a result, the velocity and magnetic field gradients for Alfvenic turbulence trace the local direction of magnetic field. In compressible MHD turbulence, numerical simulations confirmed that there exist three weakly interacting cascade modes: Alfven, slow, and fast modes (see \citealt{Cho2002CompressiblePlasmasb}). The slow modes are passive, i.e. their scaling is imposed by the Alfven modes (GS95, \citealt{Lithwick2001CompressiblePlasmas, Cho2002CompressiblePlasmasb,CL03}). Therefore, both Alfven and slow modes, i.e. the modes that contain most of the energy of the turbulent cascade, exhibit velocity and magnetic field gradients that trace magnetic field directions. 

Formally GS95 theory provides an adequate description of turbulence for {\toreferee sub-Alfvenic} motion, i.e. corresponding to Alfven Mach number $M_A=V_L/V_A$ equal to unity. Above $V_L$ is the turbulent injection velocity at the scale $L$ and $V_A$ is the Alfven velocity. However, the the theory was  {\toreferee generalized} in LV99 for $M_A<1$ where it is shown that the relations similar to those in GS95 are true for the turbulent motions at scales the physics of the strong turbulence for scales less than $LM_A^2$. For larger scales the turbulence is weak (LV99, Gaultier et al. 2000) but the fluctuations are still perpendicular to magnetic field. 

The {\toreferee super-Alfvenic} turbulence, i.e. $M_A>1$, presents a more complicated case from the point of view of gradient studies. The motions at the scales larger than $LM_A^{-3}$ are marginally constrained by magnetic field and therefore the turbulence is essentially hydrodynamic (see Lazarian 2006). However, at scales less than the aforementioned scale the GS95 scalings are applicable. Spatial filtering of the large scale contributions can help to reveal the magnetic field structure in this case (see Lazarian \& Yuen 2018a). 

The theory of synchrotron polarization fluctuations, formulated in LP16, predicts that the scaling laws of polarization fluctuations can be related to the statistics of underlying fluctuations in a rather non-trivial way.\footnote{The situation is somewhat similar to the fluctuations of intensity in thin velocity channels that were studied in \cite{LY18a}. These fluctuations exhibit a spectrum different from the one of the underlying velocity fluctuations \citep{2000ApJ...537..720L}.} Therefore, it is important to formulate general criteria when the gradients of measured fluctuations represent the underlying magnetic field structure.

The statistical properties of gradients reflect the statistical properties of the underlying anisotropic turbulence. There are two criteria for gradient technique (for both velocity and magnetic fluctuations) to reliably trace magnetic field structure:
\begin{enumerate}
\item Fluctuations at the small scale are anisotropic and the anisotropy (characterized by parallel and perpendicular dimensions of eddies  $l_{\parallel}$ and $l_{\bot}$ ) is in respect to the direction of magnetic field local to the eddies.
\item The contribution from small scale fluctuations dominates over that of the large scales.
\end{enumerate}
The first condition is satisfied due to the properties of the MHD turbulence that we mentioned above, i.e. velocities and magnetic field structures arising from Alfven and slow modes (see \citealt{CL03}) are aligned with the {\it local} magnetic field directions.\footnote{Sometime the problem arises in understanding of MHD turbulence as the concept of the {\it local} magnetic field was not a part of the original formulation of the GS95 theory. It was added to the theory later (LV99, \citealt{CV00,2001ApJ...554.1175M}).}

For example, consider an anisotropic scaling of the correlation function of an observed measure\footnote{For the case of structure function, one should replace $p$ by $-p$ .} $X(l_{\parallel}, l_{\bot}) \sim l_{\bot}^p(1 +(l_{\parallel}/l_{\bot})^2)^{p/2}$ and the anisotropy scaling $l_{\parallel} \sim l_{\bot}^q$. Statistically, the field-perpendicular gradients of $X$ are $X/l_{\bot} \sim l_{\bot}^{p-1}(1+l_{\bot}^{2(q-1)})^{p/2}$, where q indicates whether the gradients are maximally perpendicular ($q<1$) or parallel ($q>1$) to the local magnetic field, and $p$ determines whether the smallest scale dominates ($p<1$) or not ($p>1$). {\toreferee In the case of GS95, it corresponds to $p=1/3$ and $q=2/3$, indicating that both conditions are satisfied for the 3D local measurements.}

{\toreferee Readers should be careful that the statistics of magnetized turbulence change when the corresponding turbulent variables (e.g. density $\rho$, velocity $v$) are projected along the line of sight. the reason behind is that the local measurements of anisotropy are not available after compression along the line of sight} and {\toreferee thus} the line-of-sight averaged anisotropy of turbulence is determined by the anisotropy at the largest scale (see Esquivel \& Lazarian {\toreferee 2005}), i.e. the "observed" $q$ is zero {\toreferee after the line of sight averaging}. {\toreferee This change of $q$, however, would not violate the conditions for gradients to be maximally perpendicular to the magnetic field.} {\toreferee For the change of $p$ after projection,} it is easy to demonstrate that the 2D spectrum of turbulence obtained by projecting the fluctuations from 3D has the same spectral index of $-11/3$ (see \citealt{2000ApJ...537..720L}). The relation between the spectral slope of correlations and the slope of the turbulence power spectrum in 2D in this situation is $-11/3+2=-5/3$, where 2 is the dimensionality of the space. Therefore, the observed 2D fluctuations scale as $l_{2D}^{5/6}$ (i.e. $p=5/6$). The gradients anisotropy scales as $l_{2D}^{-1/6}$ {\toreferee (i.e. $p=1/6$)}. This means the contribution of the smallest scales is dominant in projected observables. These fluctuations are aligned with the local magnetic field and the observed measure reflects the line of sight averaged magnetic field. 

In our later sections, the anisotropy conditions are automatically satisfied since the GS95 scaling applies to our numerical work. As a result, we only have to examine the whether the smallest scale contribution dominates in the observed correlation or structure function scaling.

\section{Theoretical expectation of gradients of polarized synchrotron radiations}
\label{sec:theory}

\subsection{Polarized synchrotron radiation in the presence of Faraday rotation}
\label{subsec:spg.faraday}
To characterize the fluctuations of the synchrotron polarization, one can use different combinations of the Stokes parameters (see Lazarian \& Pogosyan 2012). In this paper, we follow the approach in LP16 and focus on the measure of the linear polarization $P$, which is
\begin{equation}
P=Q + i U,
\label{p1}
\end{equation}
where $Q$ and $U$ are the Stokes parameters. 

We consider an extended synchrotron region, where both synchrotron emission and Faraday rotation are taking place simultaneously. Special cases, e.g. the regions of synchrotron radiation being separated from those of Faraday rotation (see the analytical description in LP16) can be analyzed easily following the same treatment we have below. The polarization of the synchrotron emission at the source is characterized by the polarized intensity density $P_i({\bf X}, z)$, where $X$ is the two-dimensional plane-of-sky vector and $z$ is the distance along the line of sight. The polarized intensity detected by an observer in the direction $X$ at the wavelength $\lambda$ is given by
\begin{equation}
\label{polar1}
P_i({\bf X}, \lambda^2)=\int^{L}_0 dz P_i({\bf X}, z) e^{2i\lambda^2 \Phi(X, z)}
 \end{equation}
The region is extended up to the scale $L$ and the Faraday rotation measure $\Phi (z)$ is given by (see Brentjens \& Bruyn 2005)
 \begin{equation}
 \Phi (z)=0.81 \int^z_0 n_e (z') H_z(z') dz' {\rm rad~m^{-2}}
 \label{farad1}
 \end{equation}
where $H_z$ is the strength of the line of sight component of magnetic field in $\mu G$ and the distance  $z$ is measured in parsecs.
 
For our present paper, following the convention in LP16, we ignore the wavelength dependences of synchrotron polarization $P_i$ arising from the cosmic ray spectrum. This can be accomplished, for example, by determining the wavelength scaling from the intensity measurements. Thus in what follows, $P_i ({\bf X}, z)$ at the source is treated as wavelength-independent, while the observed polarization $P({\bf X}, \lambda^2)$ contains the wavelength dependence that arises from the Faraday rotation effect.
 
The Stokes parameters at the source (see Eq. (\ref{polar1})) depend on the the cosmic ray index $\gamma$. However, \cite{LP12} shows that only the amplitudes of Stokes parameters are scaled up with respect to the cosmic ray index and the spatial variation of the Stokes parameters are similar to the case $\gamma=2$.  Therefore, for our gradient study, which deals with the spatial variations of magnetic fields, it is sufficient to study only the $\gamma=2$ case,  which means that the synchrotron emissivity is proportional to squared component of magnetic field perpendicular to the line of sight.  By expressing the $Q$ and $U$ Stokes parameters at the source, we obtain
 \begin{equation}
 Q({\bf X}, z)\propto p n_e (H_{x}^2(z)-H_{y}^2(z)) ,
 \end{equation}
 \begin{equation}
 U({\bf X}, z) \propto p n_e 2H_x (z) H_y(z),
 \end{equation}
 where $p$ is the polarization fraction, which is to be assumed constant, and $n_e$ is the relativistic electron density. The definitions of the Stokes parameters above correspond to the synchrotron intensity at the source $ I({\bf X}, z)\propto H_x^2(z) + H_y^2(z).$

\subsection{Pictorial description of polarization gradients}
\label{subsec:pictorial}
\begin{figure}[t]
\centering
\includegraphics[width=0.48\textwidth]{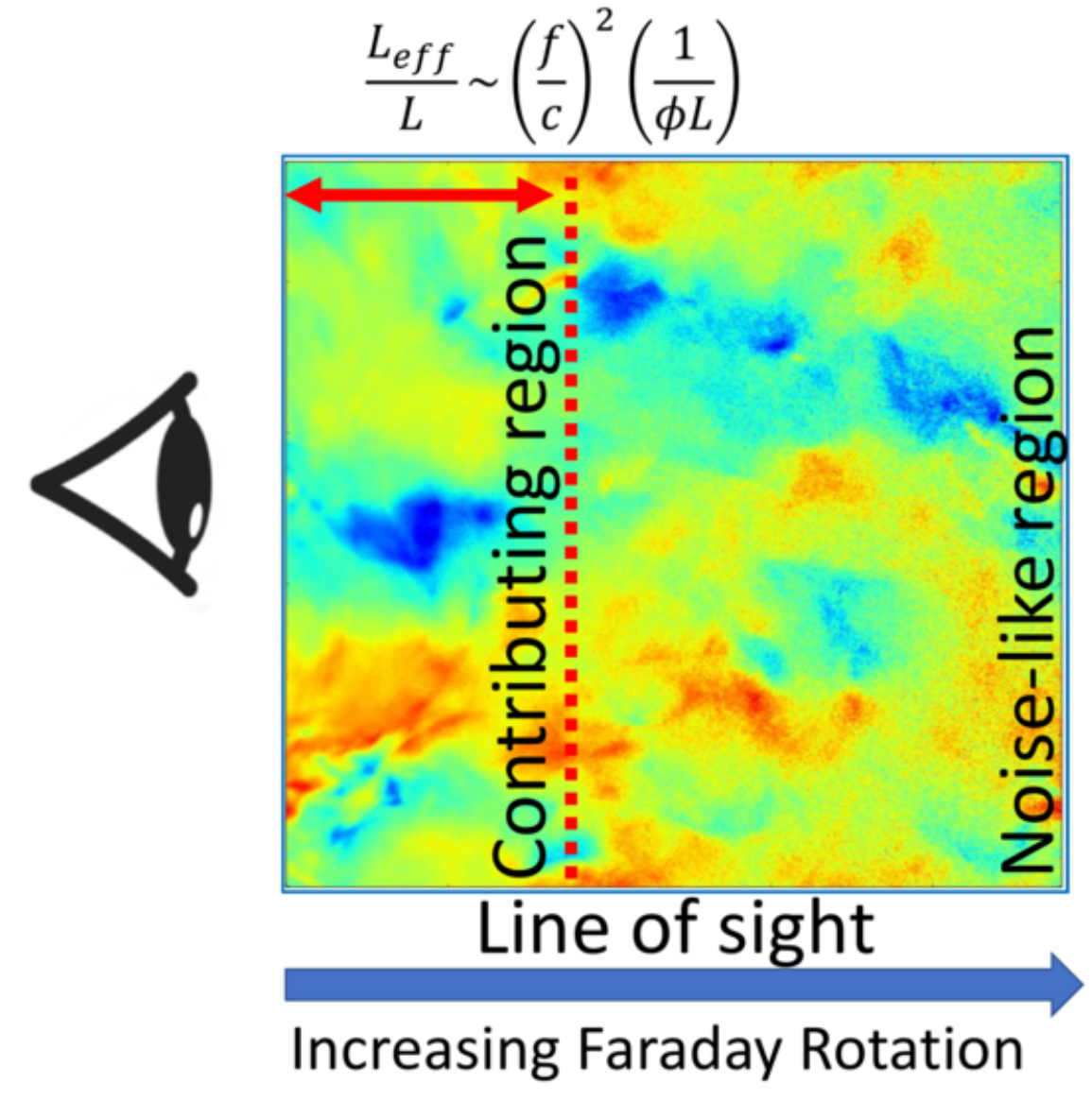}
\caption{\label{fig:illus1} A pictorial illustration showing how the effective lengthscale is defined. In the case of random magnetic field the square root of dispersion enters the expression of $L_{eff}$ instead of $B_{LOS}$ (see also Eq. (\ref{eq:el}).)}
\end{figure}

The synchrotron intensity gradients (SIGs) that we introduced in Lazarian et al. (2017) have a lot of similarities with the polarization gradients that we deal with in this paper. The most important effect that the polarization brings is related to the ability to use the effect of Faraday depolarization to constrain the region that is being sampled (see Figure \ref{fig:illus1}). The Faraday depolarization depth is defined in LP12 for both the turbulent and regular magnetic field. An additional advantage of the gradients of polarization is to sample the magnetic field structure with the Faraday rotation fluctuations. This information is complementary to what can be obtained with the SIGs and, combined with the Faraday depolarization effect, it provides the ability to restore the 3D magnetic field structure.

The three gradients we investigate in the current work are the synchrotron polarization gradients (SPGs, \S \ref{sec:limiting}) and synchrotron polarization derivative gradients (SPDGs, \S \ref{sec:FRG}). We also briefly discuss {\toreferee an variant of SPDG which is called } Faraday Rotation Gradients (FRG, \S \ref{sec:FRG}). 

LP16 describes how the statistics of fluctuations of the polarization $P$ and its wavelength derivative $dP/d\lambda^2$ depends on fluctuations of magnetic fields perpendicular to the line of sight and the Faraday effect. We use the LP16 relations to explore when the gradients measure the local magnetic field as well as to define the extend of the regions over which we collect polarized signal. 

The SPGs contains both the contributions from $Q$ and $U$. In LP16,  the discussion on the statistics of complex cross-correlation functions is limited to its real part  $\Re(\langle P(r) P^* (r+r')\rangle) = \langle Q(r)Q(r+r')+U(r)U(r+r')\rangle$.  In this work, we  also study the contributions of both $Q$ and $U$, and their quadratic combination, i.e. the phase-independent {\it polarized-intensity} $P=(Q^2+U^2)^{1/2}$ which can be studied similarly to SIGs. In \S \ref{subsec:QUcom}, we  will briefly discuss the properties of the cross-intensity $X=QU$, which is similar to the imaginary part of the cross-correlation function. 

We also explore fluctuations arising from the Faraday rotation effect in this paper. They are proportional (see Eq. \ref{farad1}) to the line of sight integral of fluctuations of $n_{e,T} B_{\|}$, where $n_{e,T}$ is thermal electron density and $B_{\|}$ is the line of sight component of the magnetic field. Mathematically, the product $n_{e,T} B_{\|}$ behaves similar to the density times the parallel component of velocity, i.e. $\rho v_{\|}$, that enters the expressions for velocity centroids (see \citealt{EL05}). Our study of velocity centroid gradients (VCGs) in GL17 and YL17ab demonstrated that they are perpendicular to the local magnetic field. Thus, we expect the FRGs to behave similarly.

Compared to the SIGs, the most promising effect related to the SPGs and SPDGs is the Faraday depolarization,  i.e. only the regions close to the observer contribute to the measured polarization, while unpolarized radiation is coming from more distant regions. This effect, analytically studied in LP16, provides the wavelength-dependent line of sight depth for probing magnetic fields (See \S \ref{sec:recipe3d}), which makes it possible to employ polarization gradients to study the 3D magnetic field structure. Fig \ref{fig:illus1} illustrates how the effective length scale is defined pictorially. Consider the situation that the observer records the synchrotron emission from the volume on left and its line-of-sight is parallel to the horizontal axis. We only study the case where the effective depth $L_{eff}$,  i.e. distance whose polarized radiation is collected effectively, is less than the total thickness $L$ of the emitting volume, i.e. $L_{eff}/L<1$. In other words, only a part of the volume close to the observer contributes to the measured Q and U. The emission coming from distances larger than $L_{eff}$ from the observer is depolarized, as we illustrated in Figure \ref{fig:illus1}. We refer to $L_{eff}$ as the {\it effective length scale} over which we probe the magnetic field with polarized synchrotron.

\subsection{Effect of density fluctuations}

One should keep in mind that density is an indirect measure of the turbulence statistics. For example, the density structures can be distorted by shocks, which can make the interpretation of density gradients in terms of magnetic field rather ambiguous (YL17b). Therefore, it is important to understand whether the fluctuations arising from the Faraday effect or from synchrotron emissivity dominate the fluctuations of polarized intensities.  In most cases, the fluctuations of synchrotron emissivity are dominated by magnetic field fluctuations because the fluctuations of relativistic electron density are negligible at small scales.  At the same time, the fluctuations of Faraday rotation are affected by both the thermal electron density as well as magnetic field fluctuations. The study in LP16 provides us with the theoretical guidance when one or the other effects dominates the fluctuations. 

\section{Numerical Simulations Employed for Testing Our Predictions}
\label{sec:numerics}

\begin{table}[t]
 \centering
 \label{tab:simulationparameters}
 \begin{tabular}{c c c c}
Model & $M_s$ & $M_A$ & $\beta=2(\frac{M_A}{M_s})^2$\\ \hline \hline
Ms0.2Ma0.02 & 0.2 & 0.02 & 0.02 \\
Ms0.4Ma0.04 & 0.4 & 0.04 & 0.02 \\
Ms0.8Ma0.08 & 0.8 & 0.08 & 0.02 \\
Ms1.6Ma0.16 & 1.6 & 0.16 & 0.02 \\
Ms3.2Ma0.32 & 3.2 & 0.32 & 0.02 \\
Ms6.4Ma0.64 & 6.4 & 0.64 & 0.02 \\ \hline
Ms0.2Ma0.07 & 0.2 & 0.07 & 0.22 \\
Ms0.4Ma0.13 & 0.4 & 0.13 & 0.22\\
Ms0.8Ma0.26 & 0.8 & 0.26 & 0.22\\
Ms1.6Ma0.53 & 1.6 & 0.53 & 0.22\\\hline
Ms0.2Ma0.2 & 0.2 & 0.2 & 2 \\
Ms0.4Ma0.4 & 0.4 & 0.4 & 2 \\
Ms0.8Ma0.8 & 0.8 & 0.8 & 2 \\\hline
Ms0.13Ma0.4 & 0.13 & 0.4 & 18 \\
Ms0.20Ma0.66 & 0.20 & 0.66 & 18 \\
Ms0.26Ma0.8 & 0.26 & 0.8 & 18 \\\hline
Ms0.04Ma0.4 & 0.04 & 0.4 & 200 \\
Ms0.08Ma0.8 & 0.08 & 0.8 & 200 \\
Ms0.2Ma2.0 & 0.2 & 2.0 & 200\\\hline \hline
\label{tt1}
\end{tabular}
 \caption {Simulations used in our current work. The magnetic criticality $\Phi = 2 \pi G^{1/2} \rho L/B$ is set to be 2 for all simulation data. Resolution of them are all $480^3$}
\end{table}
\subsection{MHD turbulence simulations }

\begin{figure}[t]
\centering
\includegraphics[width=0.49\textwidth]{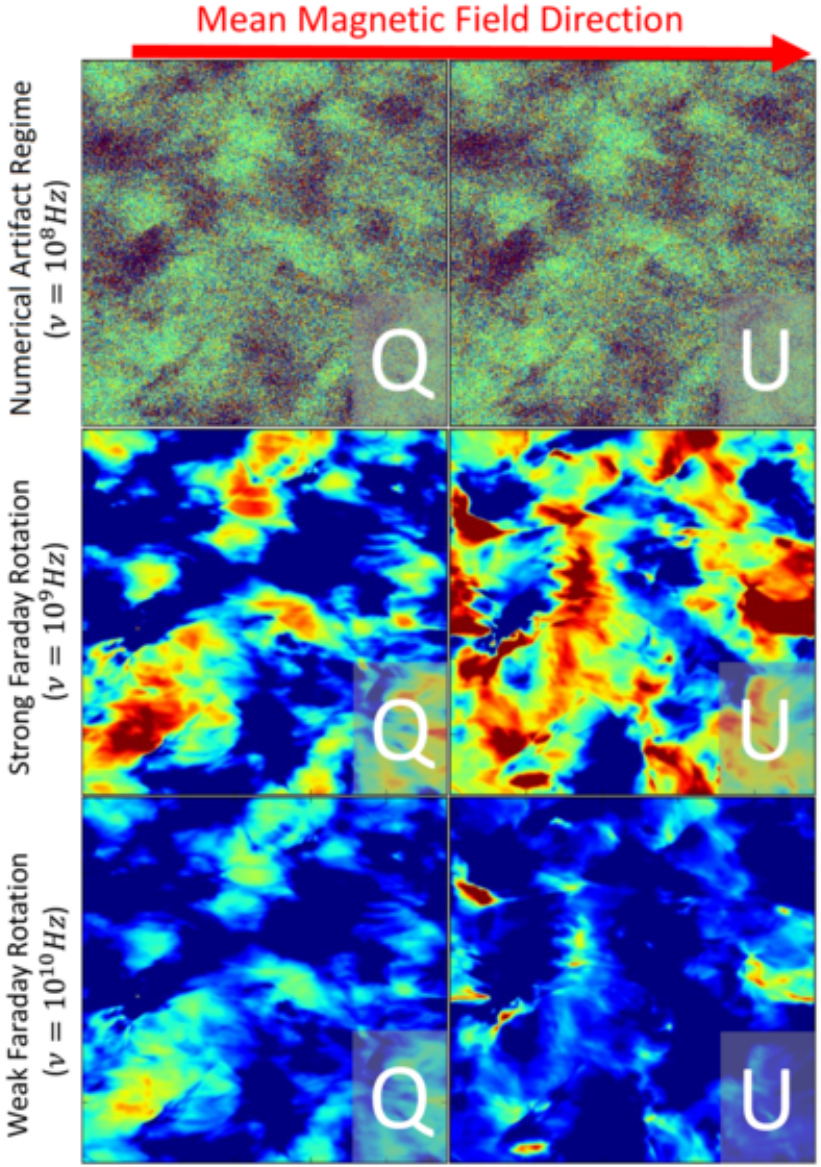}
\caption{\label{fig:QUillus} An illustration of Q and U maps in different frequencies.  }
\end{figure}

The numerical 3D MHD simulations of turbulence are performed using a single fluid, operator-split, staggered grid MHD Eulerian code ZEUS-MP/3D \citep{2006ApJS..165..188H} to set up a three-dimensional, uniform, isothermal turbulent medium.  We use a range Alfvenic and sonic Mach numbers, i.e. $M_A=V_L/V_A$ and  $M_s=V_L/V_s$, where $V_L$ is the injection velocity, while $V_A$ and $V_s$ are the Alfven and sonic velocities, respectively. The numerical parameters are listed in Table \ref{tt1} in sequence of ascending values of media magnetization given by $\beta$. The domain $M_A<M_s$ corresponds to the simulations with magnetic pressure larger than the thermal pressure, i.e.  plasma $\beta/2=V_s^2/V_A^2<1$, while the domain $M_A>M_s$ corresponds to the  $\beta/2>1$. For instance, $Ms0.4Ma0.04$ corresponds to $M_s\approx0.4$ and $M_A\approx0.04$. In this study we focus on sub- and trans-Alfvenic cases  and leave the discussion of super-Alfvenic simulations for our future publications. 

\subsection{Calculating synchrotron polarization, gradients and alignment measure}

{\bf Synchrotron radiation}.  In this work, we are interested in scalings and do not keep the numerical pre-factors. In \S \ref{sec:limiting} we investigate multi-frequency synchrotron maps, namely Position-Position-Frequency (PPF) cubes (see LP16). 

{\bf Block averaging for gradient calculations}. Gradients of polarization are calculated by taking the values of polarization in the neighboring points and dividing them over the distances between the points following the recipe of YL17a.  In this work, we focus on the smallest scale contribution (See \S \ref{sec:principle}) as we did in LY18.\footnote{In general, different scales carry information about different physical processes. Therefore gradients calculated at intermediate scales can contain important complementary information. If one is interested in gradients on larger scales than filtering of the contributions from the small scales is appropriate.} YL17b provided the explicit expressions for block-averaging and error-estimating procedures, which we will follow in this paper.

{\bf Alignment Measure (AM)}. To quantify how good two vector fields are aligned, we employed the {\it alignment measure} that is introduced in analogy with the grain alignment studies (see \citealt{2007JQSRT.106..225L}):
\begin{equation}
AM=2\langle\cos^2\theta_r\rangle-1,
\end{equation}
(see GL17, YL17a) with a range of $[-1,1]$ measuring the relative alignment between {\it rotated} gradients and magnetic fields, where $\theta_r$ is the relative angle between any two vector fields. A perfect alignment gives $AM=1$,  whereas random orientations generate $AM=0$ . We shall use $AM$ to quantify the alignments of polarization gradients in respect to magnetic field.

\section{SPGs: Random magnetic field along the LOS}
\label{sec:limiting}

\begin{figure*}[t]
\centering
\includegraphics[width=0.98\textwidth]{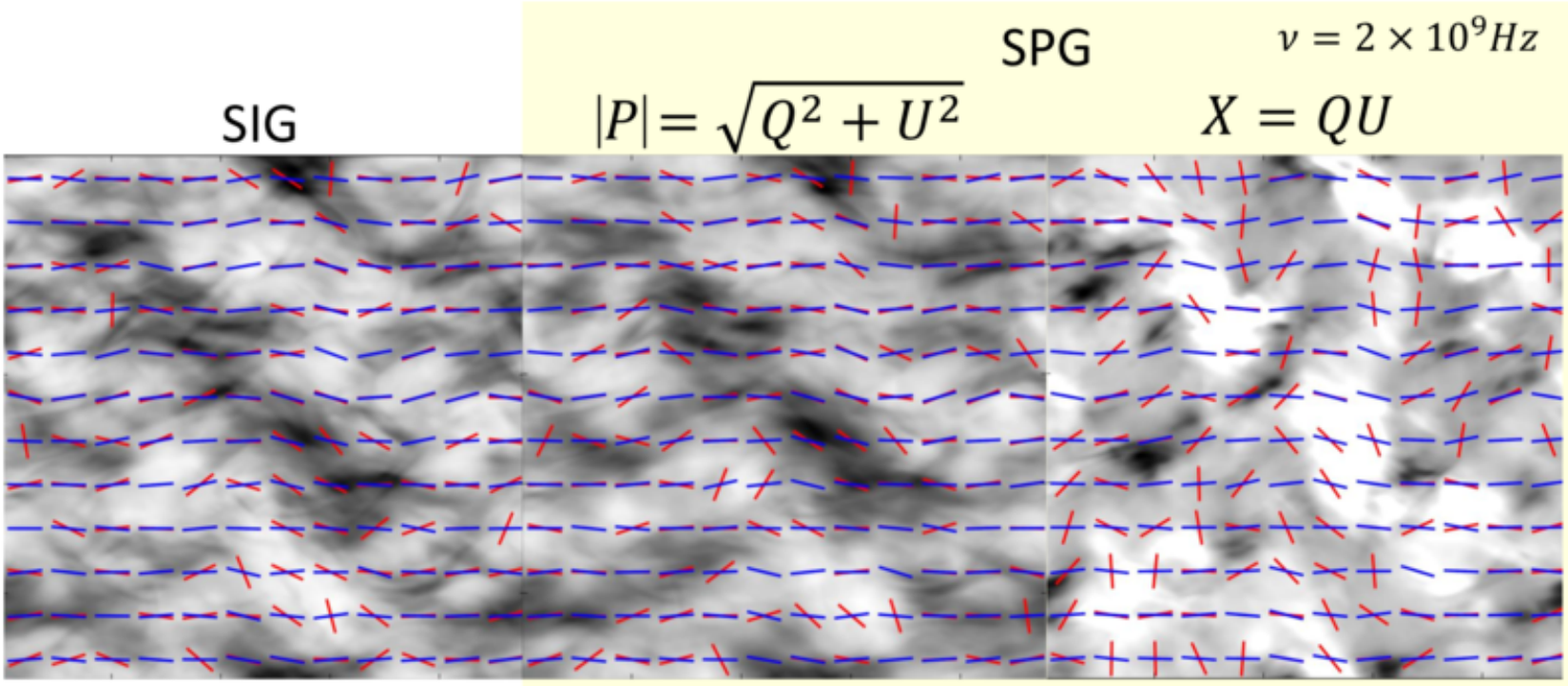}
\caption{\label{f1} (Left) The SIGs (rotated $90^o$) are shown in red versus projected magnetic field in a synthetic observation obtained with MHD simulations Ms0.08Ma0.8. (Middle and Right) The Synchrotron Polarization Gradients (SPGs, shown in red, rotated $90^o$) versus projected magnetic field (blue). }
\end{figure*}

For our studies the invariant quadratic measure  $|P| = \sqrt{Q^2+U^2}$ is used (See \S \ref{subsec:spg_weak}). However, to gain intuition with the properties of polarized synchrotron, we explore the properties of $Q$ and $U$ Stokes components. To get a theoretical guidance for our present study we will use the analytical calculations of correlation functions of polarized radiation in LP16. Notice that gradients are not equivalent to the correlation function, but statistically there is a  correspondence between the structure functions and the gradients (\S \ref{sec:principle}). When comparing the direction of gradients with respect to magnetic field, we would consistently refer to LP16 (As did in \S \ref{subsec:spg_weak}, \S \ref{subsec:spg_int} and also \S \ref{subsec:spg_strong} in the Appendix) . The various parameters that the problem depends on are listed in  Table \ref{tab:parameters}.

\begin{table*}[t]
\centering
\begin{tabular}{l l l c} 
Parameter&& Meaning & First Appearance \\\hline\hline
Scales:& & & \\
&$\lambda$& wavelength of observations &  Eq (\ref{eq:el})\\
& $\nu$& frequency of the observation & \S \ref{subsec:QUcom}\\
&$ L $ & line-of-sight extent of the emitting region & Eq (\ref{eq:el})\\
&$ R $ & separation between lines-of-sight & Eq. (\ref{weakF}) \\
&$ r_\phi $ & correlation length for Faraday Rotation Measure density & \S \ref{subsec:spg_int} \\
&$ r_i $ & correlation length for polarization at the source & Eq (\ref{strong1})  \\
&$ {\cal L}_{\bar\phi} $ & distance of one \textit{rad} revolution by random Faraday rotation & \S \ref{subsec:spg_int} \\
&$ {\cal L}_{\sigma_\phi} $ & distance of one \textit{rad} revolution by mean Faraday rotation & \S \ref{subsec:spg_weak} \\
&$ {\cal L}_{eff} $ & the smallest of ${\cal L}_{\sigma_\phi} $ and $ {\cal L}_{\bar\phi} $  &  Eq (\ref{eq:el})  \\
Spectral indexes:& & & \\
&$m_\phi$& Correlation index for Faraday RM density & Eq. (\ref{weakF}) \\
&$m$& Correlation index for polarization at the source & Eq (\ref{strong1})  \\
Basic statistical:& & & \\
&$\bar\phi$& Mean Faraday RM density & Eq. (\ref{polar1}) \\
&$\sigma_\phi$& Root Mean Squared Faraday RM density fluctuation & Eq. (\ref{eq:sigma_phi}) \\
&$\bar P_i$& Mean polarization at the source & Eq. (\ref{eq:mean_phi}) \\
&$\sigma_i$& Root Mean Squared polarization fluctuation at the source &  Eq. (\ref{weakF}) \\\hline\hline
\end{tabular}
\caption{Parameters that the synchrotron polarization statistics depends upon. Modified from LP16}
\label{tab:parameters}
\end{table*}

Before analyzing the SPGs quantitatively, we would like to list various regimes of synchrotron polarization study described analytically in LP16.  In the case of sub-Alfvenic turbulence, the source term $P_i$ is dominated by the mean field rather than the fluctuating one. The two regimes: (1) strong and (2) weak Faraday Rotation depend on whether the ratio of the scale that is sampled by polarization to the size of the emitting region, i.e.$L_{eff}/L$,  is smaller (strong) or larger (weak) than unity:
\begin{equation}
\label{eq:el}
\frac{L_{eff}}{L} \sim \frac{1}{\lambda^2L} \frac{1}{\phi}
\end{equation}
 where $\phi= max(\sqrt{2} \sigma_\phi,\bar{\phi})$ with $\sigma_\phi$ is the dispersion of random magnetic field:
\begin{equation}
\begin{aligned}
& \sigma^2_{\phi} \equiv \left\langle \Delta(n_e H_z)^2 \right\rangle \\
&= {\overline{H_z}}^2 \left\langle (\Delta n_e)^2\right\rangle
+{\overline{n_e}}^2 \left\langle (\Delta H_z)^2\right\rangle
+\left\langle \Delta n_e^2\right\rangle 
\left\langle \Delta H_z^2\right\rangle
\label{eq:sigma_phi}
\end{aligned}
\end{equation}
 with $\bar{\phi}$ is the mean Faraday rotation measure density:
\begin{equation}
\begin{aligned}
&\bar \phi \propto \left\langle n_e H_z \right\rangle = \overline{n}_e \overline{H}_z
\end{aligned}
\label{eq:mean_phi}
\end{equation}
where $\Delta n_{e,T}$ and $\Delta H_z$ are the fluctuations of the  thermal electron density and the magnetic field.

Note that $L_{eff}$ is a frequency dependent parameter (Eq \ref{eq:el}). If we have measurements from two different frequencies, we would have the Stokes parameters containing information of magnetic field directions from different line-of-sight depths. The differences between these parameters will therefore provide a measure of magnetic field between the two depths. As a result, if multiple frequency measurements are available, a 3D magnetic field distribution can be revealed by repeatedly calculating the difference of Stokes parameters obtained with the neighboring frequency measurements.

\begin{figure}[t]
\centering
\includegraphics[width=0.49\textwidth]{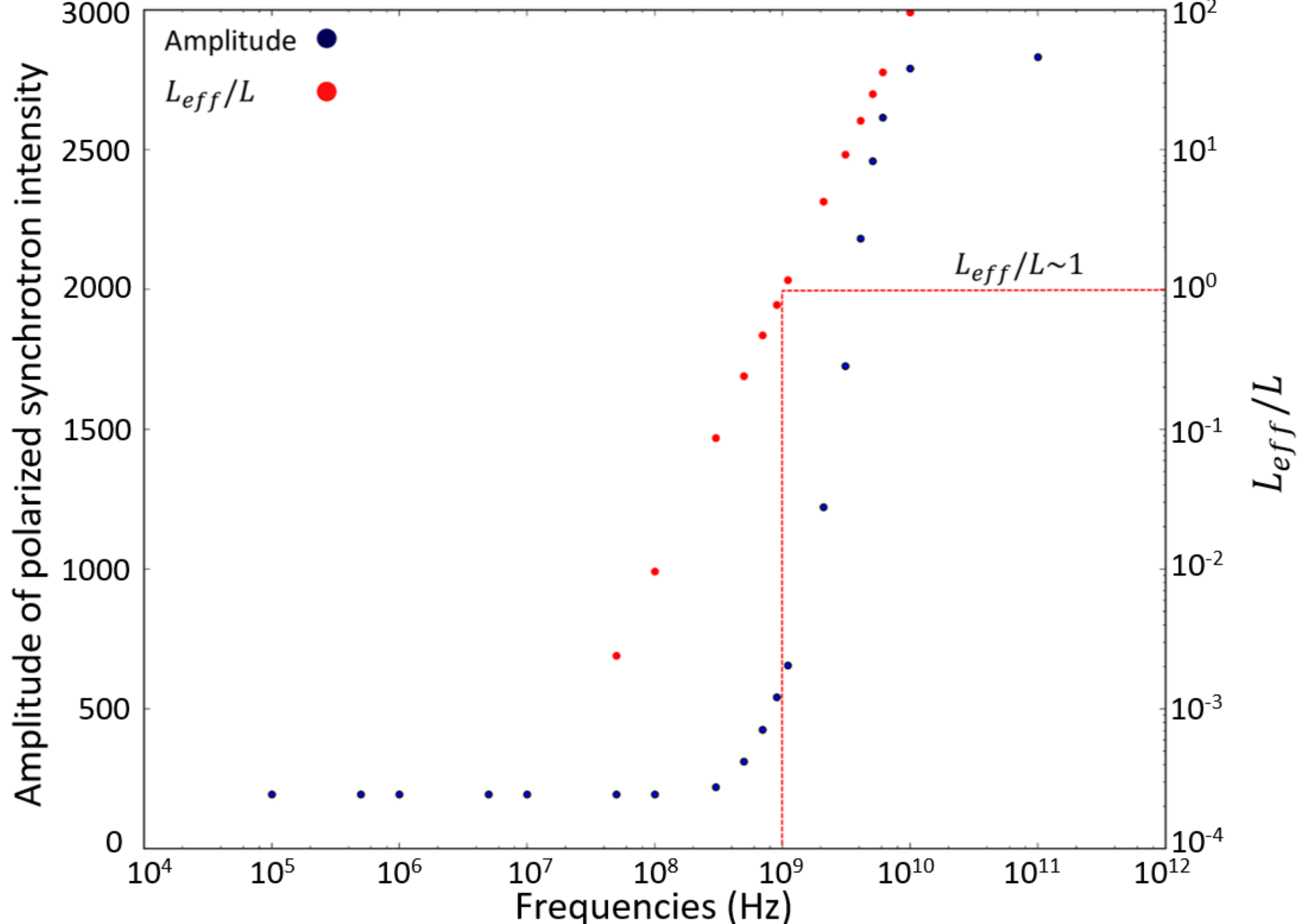}
\caption{\label{fig:amp} The variation of amplitude of $P$ and $L_{eff}/L$ versus the frequency $\nu$.}
\end{figure}

We would also like to address how the change of $|P|$ would depend on the regimes that we are going to study. To briefly characterize this, we plot the $|P|$ and $L_{eff}/L$ versus the frequency $\nu$ in Fig \ref{fig:amp}. We see that in the regime of $L_{eff}/L\sim 1$, $|P|$ changes the most. It is also interesting that when the frequency $\nu$ drops below some value, $|P|$ remains constant and significantly above zero. We recognize this as a numerical artifact of our cubes having finite resolution. We explain in \S \ref{subsec:spg_strong} how this effect arises from the finite number of points along the line of sight. {\toreferee Readers should note that our simulation resolution restricts us to synthesize ultra-low synchrotron emissions. Therefore we only investigate synchrotron emission structures in the frequency range of $100$ Mhz – $100$ Ghz. e.g. In Fig \ref{fig:amp} one can see the amplitude of polarized synchrotron emission is constant (corresponding to the numerical noise regime discussed in the appendix) when frequency is lower than $100$Mhz. 	However, the central idea of our method is to utilize the effective length $L_{eff}	$ (Eq. \ref{eq:el} to estimate the magnetic field directions for a slice in a certain LOS depth. {\torefereetwo Since the nature of the Faraday rotation effect} is the same for both low and high radio frequencies, we expect the same technique rooted in Lazarian \& Pogosyan (2016) analytical study can work for low radio frequencies.

%
}

\subsection{Effect of telescope resolution}
Based on the theory in LP16, the following equation is the {\toreferee constraint} of the telescope resolution $\Delta R$ if one wants to acquire the correct and unbiased statistics of synchrotron polarization\footnote{For any practical studies these are the scales determined by the telescope resolution. However, one may intentionally degrade the resolution to study the physical processes at different scales.} :
\begin{equation}
\label{eq:teleres}
1/r_i<\lambda^2 \mathrm{max}(\sigma_{\phi},\bar\phi)  < 1/\Delta R,
\end{equation}
where $r_i$ is the correlation length for polarization, $\sigma_{\phi}$ and $\bar\phi$ characterize the Faraday rotation effect for random and regular magnetic field, respectively (see more explanation in Table 2). Naturally, this also sets a limitation for the gradient technique. The same criterion also applies to the numerical analysis we perform in the present paper. For instance, numerical artifact {\toreferee emerges} when the inequality (Eq (\ref{eq:teleres})) is violated (See \S \ref{subsec:spg_strong}).

\subsection{SPGs: Competition between Q and U}
\label{subsec:QUcom}

We would like to give a visual illustration on how the Faraday rotation from the strong and weak regimes change the structure of both $Q$ and $U$. Fig. \ref{fig:QUillus} shows the structures of $Q$ and $U$ map {\it in relative scales} for three different regimes of Faraday Rotation from our cube Ms1.6Ma0.53, which is a sub-Alfvenic cube. The dispersion for the quantity $\sigma_\phi$ is 1.42 in numerical cube Ms1.6Ma0.53. We select three frequencies $\nu=100MHz, 1GHz, 10Ghz$  for our case study, which corresponds to $L_{eff}/L \sim0.0068,0.68,68$, respectively. For this cube, the mean magnetic field is pointing to the right and has minimal oscillations vertically. Therefore we expect a dominance of Q over U. For the best presentation of Fig \ref{fig:QUillus}, we adjusted the color bar to relative scales so that most structures are seen visually. In Fig \ref{fig:QUillus} colors are adjusted in a way to show the maximal amount of visual information. 

As gradients identify  the orientation of intensity structures, we show {\toreferee that} these structures first in Fig. \ref{fig:QUillus}. One can see that in case of $\nu=100Mhz$ both Q and U maps are being dominated by noise. We recognize this as a numerical artifact (See \S \ref{subsec:spg_strong}). As the frequency increases, the noise-like structures are less prominent. For our case, the structures of Q maps are aligned with the mean magnetic field direction while that for U tend to be perpendicular to magnetic field. We shall see how the competing effects between Q and U in $|P|=\sqrt{Q^2+U^2}$  change the gradient orientation. We also reported the ratio between $\langle U\rangle$ and $\langle Q \rangle$ with respect to frequency. In fact, $\langle U\rangle/\langle Q \rangle \sim$ is 1 when $\nu=1GHz$ ,whereas is $\sim 0.016$  when $\nu=10GHz$. Note that the properties of  $|P|=(Q^2+U^2)^{1/2}$ and $X=QU$  are related to the relative importance of $Q$ and $U$.

\subsection{SPGs for weak Faraday Rotation effect}
\label{subsec:spg_weak}

We start with the simple case of weak Faraday rotation. This situation corresponds to the case when the wavelength of radiation is sufficiently short so that Faraday rotation can be disregarded.  In this situation, the mathematical structure of Eq. (\ref{polar1}) is similar to that of unpolarized synchrotron intensities. Therefore it is not surprising that we get the structure of the SPGs similar to the structure of the SIGs. 

The theoretical expectation of LP16 suggests that the weak Faraday rotation, i.e. when the correlation scale of the synchrotron emission correlation\footnote{This is essentially the correlation scale of magnetic field fluctuations.} is less than the depolarization scale, i.e. $r_i < {\cal L}_{\sigma_\phi}$, has a only a weak effect on the correlation of observed polarization.  As we cannot obtain $r_i$ beforehand, we enforce our weak rotation case to be $L< {\cal L}_{\sigma_\phi}$. This ensures the ratio in Eq (\ref{eq:el}) is greater than unity and $r_i < {\cal L}_{\sigma_\phi}$.  The calculations in LP16 provides:
\begin{equation}
\begin{aligned}
\label{weakF}
&\left\langle \left| P({\bf X}_1) - P({\bf X}_2)\right|^2 \right\rangle  \propto \sigma_i^2  R^{1+\bar{m}} ~,\\
&\textit{where} \qquad L_{eff} > L
\end{aligned}
\end{equation}
where $ \bar{m} \equiv \mathrm{min}(1,m) ${\toreferee,$\sigma_i^2$ is the RMS polarization fluctuation at the source} and $m$ is the correlation index of the polarization at the source.  From \S \ref{sec:principle} we know that the power-law dependence of R  satisfies the expected $p<1$ requirement \footnote{\kh Correlation function should be $\sim R^{-(1+\bar{m})}$, in which $R$  has a power of $1+m$, which for the case of Kolmogorov-type scaling, i.e. $m=2/3$, provides the fluctuation $\delta b_l\sim l^{5/6}$ with the gradients $\sim l^{-1/6}$, i.e. the correlation function is the most sensitive to the fluctuations at the smallest scales following \S \ref{sec:principle}.} {\toreferee (See \S \ref{sec:principle})}.

From the principle in \S \ref{sec:principle} we confirm that gradients in the weak Faraday rotation case are able to trace the projected magnetic field. LP12 showed that the anisotropy arising from a $\sigma_i\propto H^2$ term in Eq (\ref{polar1}) can be measured with synchrotron intensities. As the spatial anisotropy part is not altered in Eq (\ref{weakF}), we predict that the direction of the SPG is similar to that of SIGs.

The prediction of the directional similarity of SPGs and SIGs in the weak Faraday rotation case is tested  by applying the block averaging recipe of YL17a to the polarization $|P|=(Q^2+U^2)^{1/2}$ as well as to $X=QU$. This is illustrated in Fig \ref{f1}, where the SIGs and the SPGs calculated for the weak rotation case ($\nu=2GHz$) are compared for the projected magnetic field for the data cube Ms0.08Ma0.8. One can already see the structural similarity of the SIGs and SPGs in Fig \ref{f1}. When Faraday Rotation is weak, the polarized intensity is dominated by the Stokes U in our synthetic cubes. Gradients from the cross-intensity $X=QU$ trace magnetic fields much worse compared to $|P|$ (see Fig \ref{f1}).

The advantage of phase-independence for $|P|$ is very useful when we reconstruct the 3D  magnetic field  distribution in \S \ref{sec:recipe3d}. The utility of cross-intensity $X$ for the gradient analysis will be explored elsewhere. Unless otherwise stated, in what follows we  deal with $|P|$.

\subsection{SPGs for strong Faraday Rotation effect}
\label{subsec:spg_int}
\begin{figure}[t]
\centering
\includegraphics[width=0.49\textwidth]{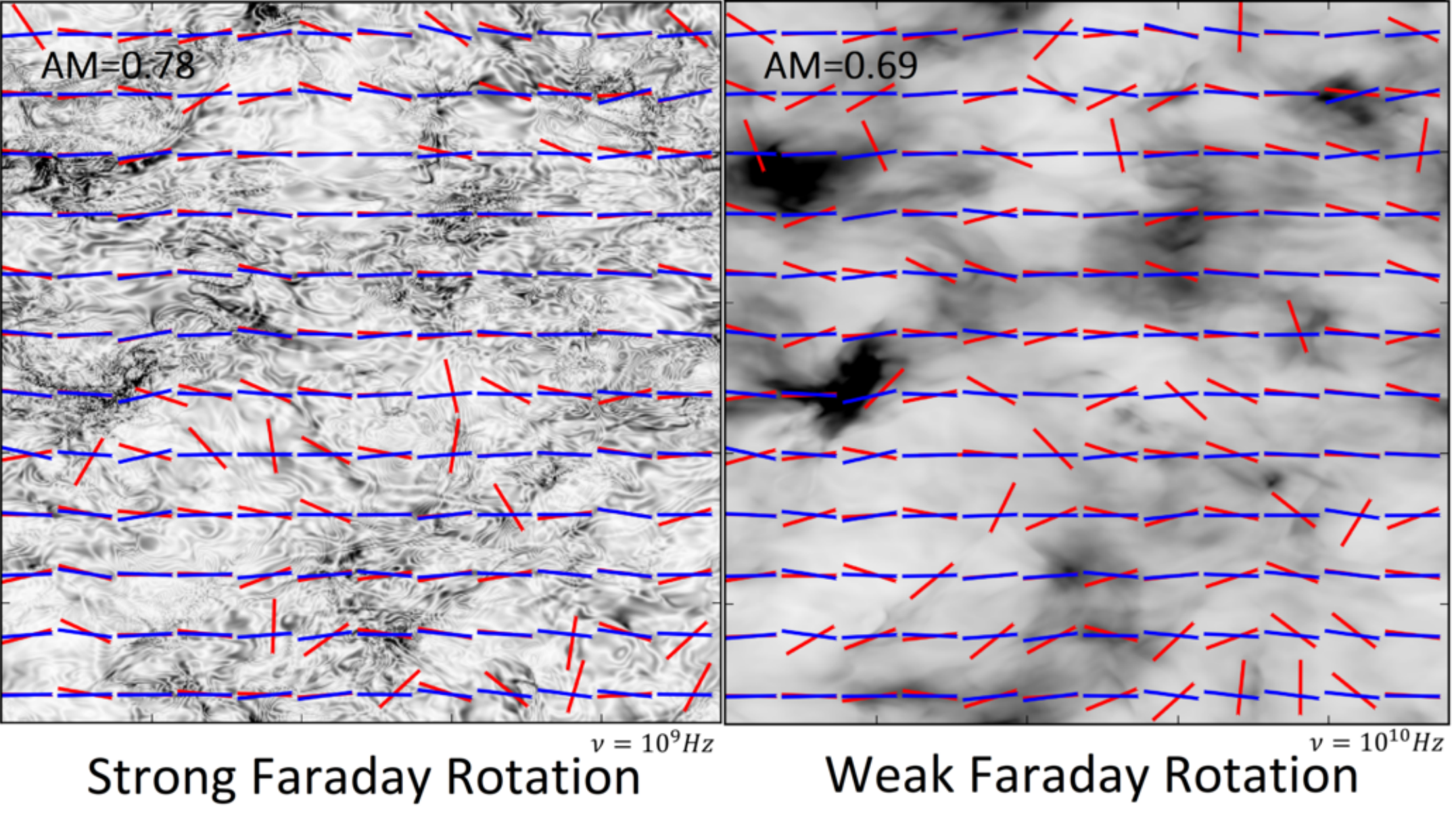}
\caption{\label{f2} The SPGs for $|P|$ (shown in red, rotated $90^o$) compared with the projected magnetic field (blue) within the MHD turbulence simulation data cube Ms1.6Ma0.53 for the weak and strong regimes of Faraday rotation.   }
\end{figure}

The strong Faraday rotation case opens a unique way for 3D magnetic field studies. Fig. \ref{f2} illustrates how the SPGs are aligned compared with the projected magnetic field. The left panel of Fig \ref{f2} illustrates the SPGs of $|P|$ in the case of strong Faraday Rotation corresponding to the  $\nu=1GHz$. On the right panel of Fig \ref{f2} we show the map with weak Faraday Rotation. One can visually see both the striations along the magnetic field lines as well as the 90 degrees rotated gradients being aligned with the magnetic field directions. 

Below we explain why the strong Faraday rotation setting is so advantageous from the point of view of tomographic studies of magnetic field. Indeed, we will show that the scales for the decorrelation of the polarized signal arising from the rotation by random (scale ${\cal L}_{\sigma_\phi}$) and regular (scale ${\cal L}_{\bar\phi}$) magnetic fields (see Table \ref{tab:parameters})

For strong rotation, we cannot replace $r_i$ by $L$ as described in \S \ref{subsec:spg_weak} because the polarization is not sampled through the entire volume. For instance, consider the dominance of turbulent Faraday decorrelation, i.e. ${\cal L}_{\sigma_\phi}<{\cal L}_{\bar\phi}$. The strong regime ${\cal L}_{\bar\phi}\ll r_i$ and therefore the polarized radiation is sampling only a small part of the nearby synchrotron eddy. The result for the polarization correlation that is obtained in LP16 for this regime is 
\begin{equation}
\begin{aligned}
&\left\langle P({\bf X}_1)P^*({\bf X}_2) \right\rangle \propto \xi_i({\bf R},0) r_i^{\widetilde{m}_\phi} R^{-(1+\widetilde{m}_\phi)}\\
& \textit{where} \quad \quad L_{eff} < L 
\label{strong1}
\end{aligned}
\end{equation}
where $\widetilde{m}_\phi=min(1, m_\phi)$. Again, from \S \ref{sec:principle} that the requirement for small-scale gradients to dominate is satisfied.  For small separations $R\ll r_\phi$, the correlation function of synchrotron emissivity at the source is Eq. (\ref{strong1})  which is nearly constant and therefore the observed fluctuations of   $|P|$ are dominated by the Faraday effect (see \S \ref{sec:LP18} for the illustration). 

At scales less than the correlation scale of Faraday rotation $r_\phi$  (see Table \ref{tab:parameters}) the gradients are dominated by the effects of Faraday rotation. As the correlation function of polarization fluctuation increases with the decrease of the scales, the gradients of the polarization fluctuations are dominated by the smallest scale fluctuations. Therefore, we expect the gradients of $P({\bf X})$ to be perpendicular to the plane of sky magnetic field projection within the distance ${\cal L}_{\bar\phi} \equiv (\lambda^2 \bar\phi)^{-1}$ (see Table \ref{tab:parameters}) from the observer.  It is important that, by changing the  wavelength of the radiation $\lambda$, one can change the thickness of the volume inside which SPGs can trace magnetic field. 

\subsection{Alignment Measure in different regimes and numerical artifacts}

\begin{figure}[t]
\centering
\includegraphics[width=0.49\textwidth]{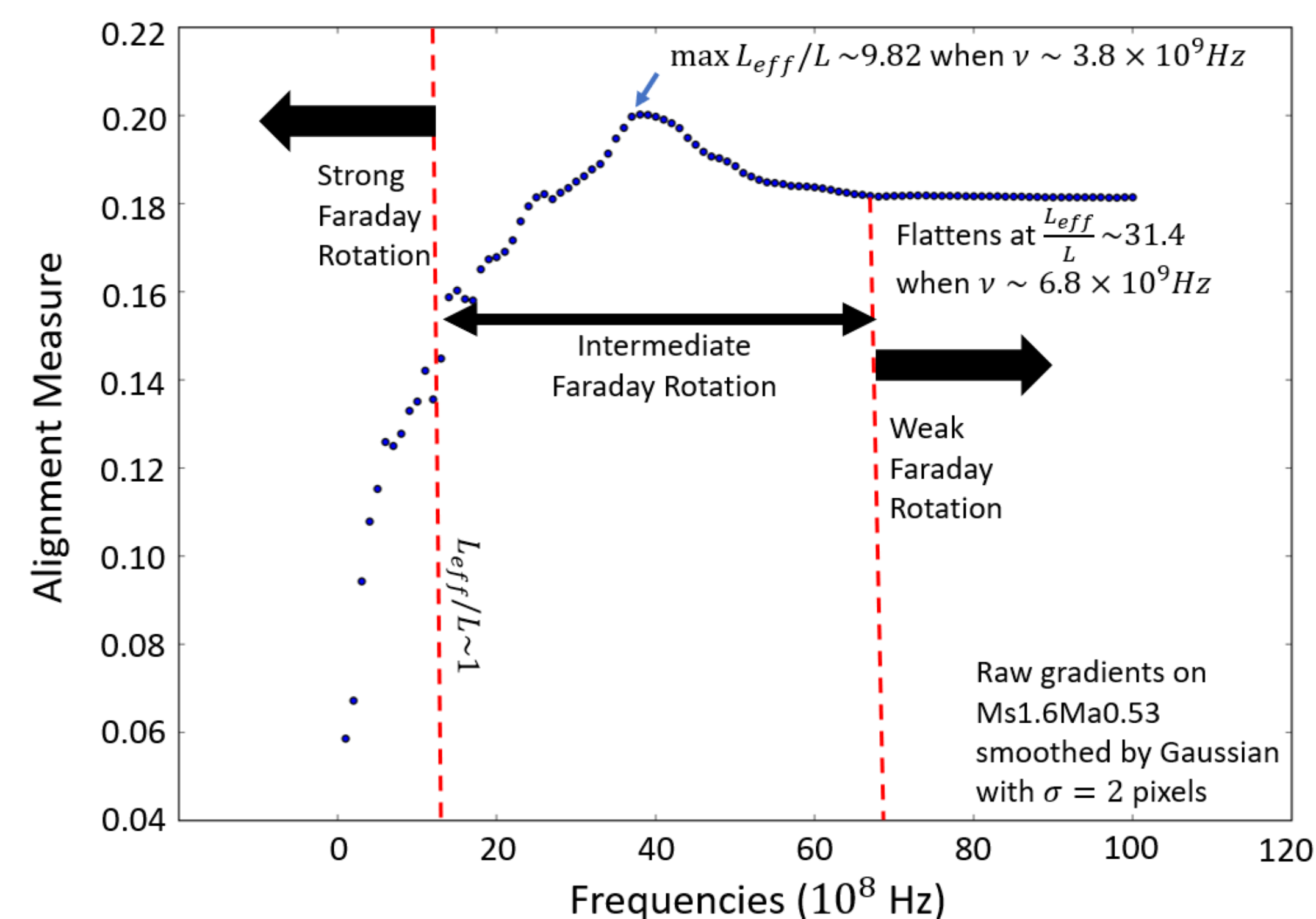}
\caption{\label{f2s} A scatter plot showing how the variation of frequencies can affect the alignment measure for the cube Ms1.6Ma0.53.}
\end{figure}

We analyze the change of AM with respect to projected magnetic field directions {\it before block averaging} for various $\nu$. Fig. \ref{f2s} shows a  plot of the alignment measure with respect to the frequencies when the gradients are smoothed by a Gaussian kernel (see LYLC) of $\sigma=2$ pixels. 

When $\nu \ge$ is 6.8 GHz the change of alignment measure is infinitesimal and shows the same features as the SIGs, which approximately corresponds to the weak Faraday Rotation regime. As the frequency decreases, the striations seen in Fig \ref{f2} get to be dominant and this brings a higher AM than that in raw SIGs with its peak at $\nu=3.8 GHz$. As the frequency continue to decrease, the AM quickly drops to zero due to the dominance of noise-like structures already seen in Fig \ref{f2}. As a result, gradients in this very  low-frequency synchrotron maps are isotropic and incapable of tracing magnetic field. We mark the approximate range for the {\toreferee two} regimes we discussed above in Fig \ref{f2s}. The peak we recognized in Fig \ref{f2s} is helpful for construction of three dimensional magnetic field, because the corresponding frequency that has {\toreferee a peak in the AM-frequency plot (Fig \ref{f2})} is related to the effective screening length $L_{eff}$.

\section{The Faraday Rotation Measure and the Synchrotron Polarization Derivative Gradients (SPDGs)}
\label{sec:FRG}

From the theoretical study in LP16,  we know that the fluctuations of $|P|$ are mostly coming from the variations of synchrotron emissivity at the source rather than the fluctuations from Faraday rotation. To study the fluctuations of Faraday rotation, LP16 proposed to use $dP/d\lambda^2$ measures. In the case  that synchrotron emission and Faraday rotation regions being spatially distinct, the aforementioned measure is directly proportional to the Faraday rotation measure $\sim \int n_e B_{\|} dz$. A important relation was found in LP16 for the spatial correlations of $dP/d\lambda^2$ ,when the volumes of synchrotron emission and Faraday rotation coincide. The relative importance of the Faraday rotation fluctuations and synchrotron emission fluctuations depends on the separation of points at which the correlation is calculated (LP16). In terms of gradients of $dP/d\lambda^2$, this means that  one can probe the anisotropy of the two types of fluctuations by calculating the gradients at different scales. The latter can be achieved by calculating the gradients not using the adjacent points, but taking the points separated by a given distance. Such a study may provide valuable additional information. For instance, the differences between the synchrotron emission and the gradients  that are influenced by the Faraday effect can  give insight to the distribution of the rotation measure.

\begin{figure}[h]
\centering
\includegraphics[width=0.48\textwidth]{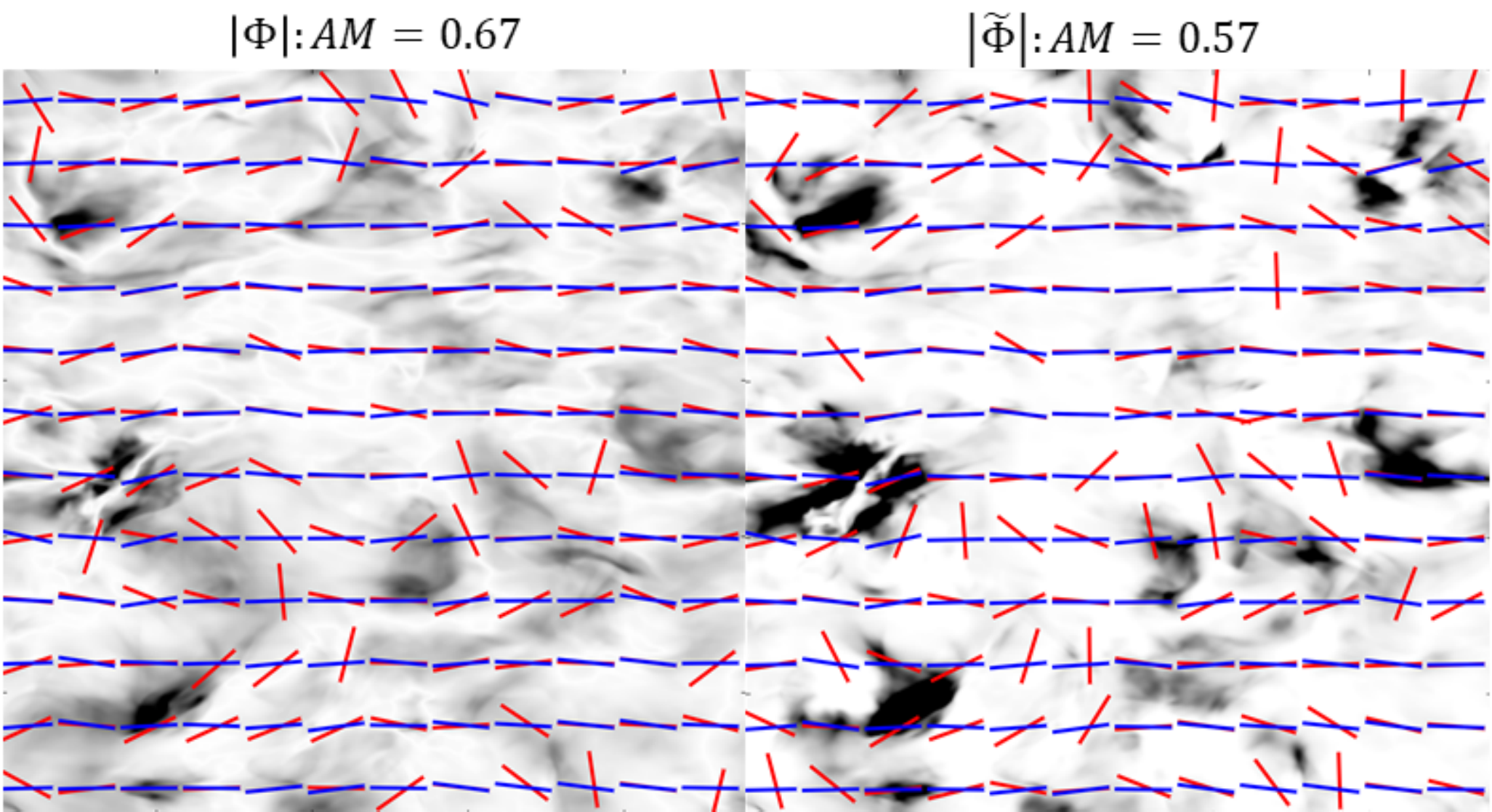}
\caption{\label{fig:frg} (Left) The Faraday Rotation Gradients (FRGs, shown in red) are turned 90 degrees to be compared with the projected magnetic field (blue) within the MHD turbulence simulation data cube. (Right) The normalized SPDGs directly coming from the PPF cube at $\nu=10GHz$ with $\Delta \nu=100MHz$.  }
\end{figure}

We start with the simplest case. The Faraday rotation measure can be available when the polarized signal passes through the Faraday rotating screen. This is a valid approximation when the synchrotron {\toreferee emission} and Faraday rotation regions are separated. As we mentioned earlier, the structure of  Eq \ref{farad1} is very similar to the Velocity Centroid, i.e. $C\propto \int dz \rho(z) v_z(z) $, where $z$ is the Line of Sight (LOS) axis. Therefore  we expect that the  Faraday Rotation Gradients (FRGs) have  similar properties to the Velocity Centroid Gradients (VCGs)  in e.g. YL17a. In left panel of Figure \ref{fig:frg} the $90^o$ rotated FRGs are compared to the projected magnetic field for the synthetic observations obtained with the cube Ms1.6Ma0.53. To compare with the weighted FRGs (see below), the background map and the gradients are both based on the absolute value of the Rotation Measure, even though the distribution of gradients is independent of signs of the map. The alignment measure for the FRGs is significant ($AM=0.67$) which means that the FRGs are capable of successfully tracing magnetic fields.\footnote{We expect that the alignment can be increased further if the additional techniques that we tested for velocity gradients, e.g. angle constraint and moving window (see LY18), are applied.}

The FRG is a limiting case of long wavelengths for the Synchrotron Polarization Derivative Gradients (SPDGs), which can be used also in situations when the Faraday fluctuations and synchrotron emission arise from the same volume. The SPDGs are gradients of the polarization derivative $dP/d\lambda^2$, whose correlations
\begin{equation}
\left\langle \frac{d P({\bf X}_1)}{d \lambda^2}
\frac{ d P^*({\bf X}_2)}{d \lambda^2} \right\rangle ,
\end{equation}
were studied in LP16. In the case of the strong mean field dominated over random field along the line of sight, LP16 predicted that 
\begin{equation}
\left\langle \frac{d P({\bf X}_1)}{d \lambda^2}
\frac{ d P^*({\bf X}_2)}{d \lambda^2} \right\rangle \propto R^{1+\widetilde{m}_\phi}
 \label{add_1}
 \end{equation}
where $\widetilde{m}_\phi=min(m, 1)$ see Table \ref{tab:parameters}. This is  suggestive that the SPDGs, i.e. the gradients of $\frac{d P({\bf X}_1)}{d \lambda^2}$ are arising from the  fluctuation of Faraday rotation: There is a power-law dependence on the correlation function (Eq. \ref{add_1}) to the Faraday Rotation Measure correlation index.

The situation is more complex in the case of weak mean magnetic field:  the correlations of intrinsic synchrotron polarization at the source $P_i$ and the fluctuations of the Faraday rotation $\Delta(n_e H_z)$  enter symmetrically in the correlation of the polarization derivatives. The dominant asymptotic contribution depends both on the correlation scale of the two types of correlations and their indexes (see LP16). Assuming that $r_\phi<r_i$ and $m_\phi < m$, the structure function of the correlation derivatives obtained in LP16 is
\begin{equation}
\begin{aligned}
&\left\langle \left| \frac{d P({\bf X}_1)}{d \lambda^2}
- \frac{ d P({\bf X}_2)}{d \lambda^2}\right|^2 \right\rangle
\propto \sigma_i^2 \sigma_\phi^2 L^3 (r_i/L)^{m} R^{1+m_\phi}/r_\phi^{m_\phi} \\
& where \quad\quad   R <  r_\phi, \quad m_\phi \le m~.
\label{add_2}
\end{aligned}
\end{equation}
 suggesting again the dominance of the Faraday fluctuations in the SPDGs.

In reality, obtaining the rotation measure in the form of Eq \ref{farad1} is usually difficult. Therefore we would be more interested in the so called {\it Polarization-weighted} Faraday Rotation Measure $\tilde{\Phi}$:
\begin{equation}
\label{eq:faradw}
\tilde{\Phi}= -i\frac{dlog(P)}{d\lambda^2} = \frac{\int^{L}_0 dz \Phi(X, z) P_i({\bf X}, z) e^{i\lambda^2 \Phi(X, z)}}{\int^{L}_0 dz P_i({\bf X}, z) e^{i\lambda^2 \Phi(X, z)}}
\end{equation}
which is simply the log-derivative of Eq (\ref{polar1}). As we described above, we would be more interested in the modulus of the structure of the weighted measure. For simplicity, we take $\lambda \to 0$ (i.e. $\nu \to \infty$) to suppress the Faraday Rotation effect. The right of Figure \ref{fig:frg} shows $\tilde{\Phi}$ for the same cube Ms1.6Ma0.53 by taking the differences on the $(Q,U)$ maps from $\nu=10GHz$ with $\Delta \nu=100MHz$  ( $\Delta \lambda^2 = c^2/f^3 \Delta f =9 \times 10^{-6}$ m), i.e. 
\begin{equation}
\begin{aligned}
\Delta Q &= Q(\lambda=10.1 GHz)- Q(\lambda=10GHz) \\
\Delta U &= U(\lambda=10.1 GHz)- U(\lambda=10GHz) \\
\end{aligned}
\end{equation}The differences of the polarized intensity is then $|\tilde{\Phi}| = \sqrt{\Delta Q^2+\Delta U^2}$. One can observe that both the structure of the map of $|\tilde{\Phi}|$ or the gradients of  $|\tilde{\Phi}|$  behave similarity to the  true Rotation Measure  (Eq (\ref{farad1})). That means the studies with high enough $\nu$   can deliver the Faraday rotation measure from the synchrotron emitting volume.
Moreover, the alignment measure of the gradients of $|\tilde{\Phi}|$  is relatively high ($AM=0.57$), which is suggestive that $|\tilde{\Phi}|$ is also a good measure of magnetic  field direction for the case of high frequency radiation (i.e. small $\lambda$).

\section{SPGs: Faraday rotation by mean field}
\label{sec:spg-meanb}

A particular regime when strong Faraday rotation is induced by  a component of the mean magnetic field along the line of sight  was studied in LP16. In this section, we explore what this regime means  to the gradient technique on tracing magnetic field.

\begin{figure}[t]
\centering
\includegraphics[width=0.48\textwidth]{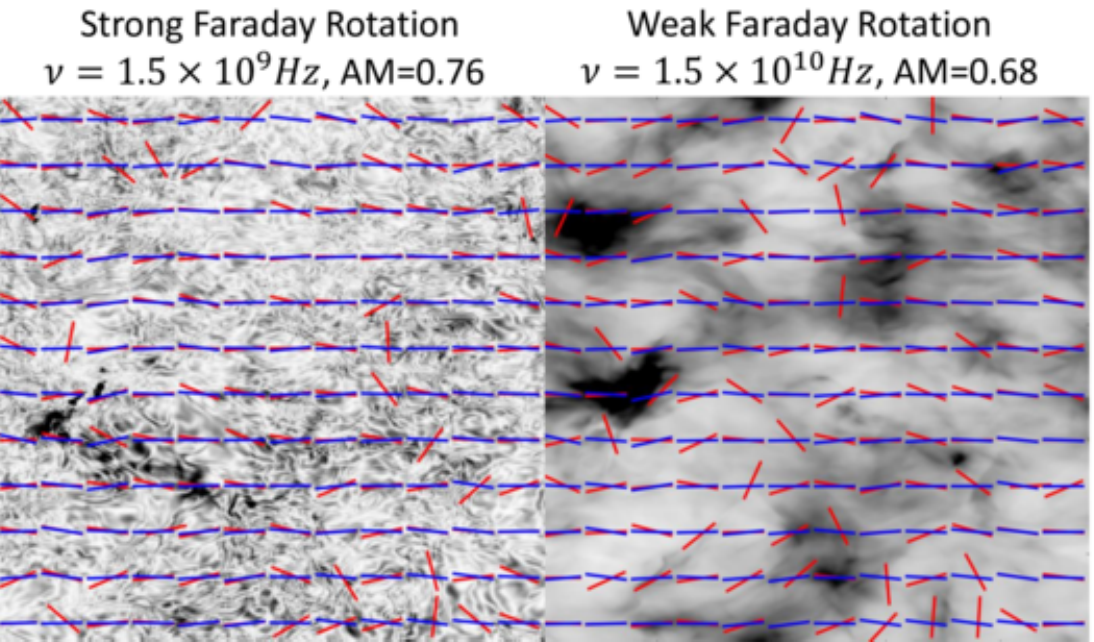}
\caption{\label{fig:meanfield}  Illustration of the mean-field Faraday Rotation effect for the two regimes. From left to right: the strong and weak field. The  SPGs ($90^o$ rotated,red) for $|P|$ are plotted against the projected magnetic field (blue) overlaid on their respective $|P|$ map. }
\end{figure}
{\kh In the strong Faraday rotation case,} if the mean field dominates the Faraday rotation ${\cal L}_{\sigma_\phi} > {\cal L}_{\bar\phi}$ (see Table \ref{tab:parameters} ),  LP16 predicts that under the condition of ${\cal L}_{\bar\phi} < r_i$, we have 
\begin{equation}
\begin{aligned}
&\left\langle P({\bf X}_1)P^*({\bf X}_2) \right\rangle \sim
L \; {\cal L}_{\bar\phi} \; \xi_i({\bf R},0)~,\\
& where \quad \quad \lambda^2 r_i \bar\phi > 1~.
\label{meanF}
\end{aligned}
\end{equation}

We test this by artificially putting a line-of-sight component of magnetic field with $\langle B_{LOS} \rangle = 1$ and $\sigma_B=0.2$. The field is  applied as an external agent and is not self-consistent with the simulations. However, it serves the purpose of illustrating the effect that we discuss in this section. The mean field component $\bar{\phi}$ is then larger than the oscillating part and is dominant in the calculation of effective length scale. We then pick two frequencies $1.5GHz$ and $15GHz$, which correspond to $L_{eff}/L=1.02$ and $102$ respectively. Fig \ref{f2s} shows the gradients in the two regimes and also the underlying intensities.

The gradient analysis of $|P|$ in this section  is similar to \S \ref{sec:limiting}, so we shall directly show the results from all three cases. Fig. \ref{fig:meanfield} illustrates the gradients for the three regimes. One can see a similar trend for alignment measure to spike when $L_{eff}/L$ is in the order of unity (i.e. $\nu =1.5GHz$). The dependence of AM on frequency is very similar for the mean field case to the one in Fig \ref{f2s} for the random field case. Therefore, we in what follows shall unify the treatment for both the random field and the mean field Faraday Rotation effect. 

Note, that even the presence of the small random field changes the amplitude in Eq. (\ref{meanF}). It is easy to see that for $L> {\cal L}_{\sigma_\phi}$, it is ${\cal L}_{\sigma_\phi}$ that should substitute $L$ in Eq. (\ref{meanF}). The change of the amplitude does not change the properties of gradients, however.

\section{Towards constructing of 3D map of magnetic fields using SPGs}
\label{sec:recipe3d}

Aside from the average magnetic field along the line of sight, the special properties of Faraday depolarization effect allow one to extract the information of magnetic field morphology along the line-of-sight and also potentially the three-dimensional structure of magnetic field. The Faraday depolarization effect limits sampling of the magnetic fields within the boundary $L_{eff}$ from the observer. In this section we illustrate the ability of the technique showing that by changing the wavelength one can trace changes of the magnetic field direction that take place along the line of sight. Much more, however, can be done with the technique and in the Appendix D we discuss the possibility of obtaining the true 3D structure of magnetic field with the polarization gradients.  

We start with summarizing the properties of SPGs that we found. First of all, in both weak and strong Faraday rotation regimes, the SPGs trace magnetic fields well. More importantly, in the strong Faraday rotation regime, the SPGs sample magnetic field up to the depth $\sim L_{eff}$. It is important that within the volume limited by $L_{eff}$, the directions traced by SPGs, unlike the polarization directions, are not distorted by the Faraday rotation effect. As $L_{eff}$ depends on the wavelength $\lambda$, by changing $\lambda$ one can sample magnetic field at different distances. As a result, the 3D distribution of the plane of sky magnetic field components  can be obtained.\footnote{Similar to the SIGs, it is possible to get the amplitude of magnetic field perpendicular to the line of sight using the synchrotron intensities. Thus, not only is the direction of the magnetic field vector, but its amplitude is also available.}

We illustrate the capabilities of the SPGs using a toy model of magnetic field distribution. The SPGs sample the {\it averaged} magnetic field direction for depths $z\in [0,L_{eff}]$. To test this numerically we set up an artificial case based on the numerical cube Ms1.6Ma0.53 by dividing it along the line of sight into three pieces. We rotate the density and magnetic field structures in the second piece for $30^o$ but keep the other two pieces unchanged. We shall refer to the altered cube as the {\it partially rotated cube}. 

\begin{figure*}[t]
\centering
\includegraphics[width=0.98\textwidth]{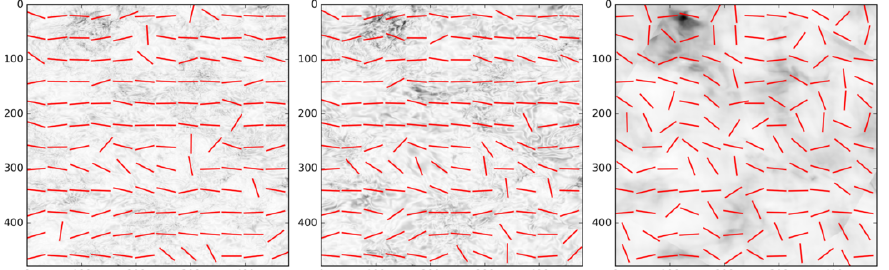}
\caption{\label{fig:spg3d-illus} Sampling of magnetic field structure along the line of sight by changing the radio frequency. From left to right: $\nu = 0.85, 1.2, 7 GHz$, which correspond to $L_{eff}/L\sim 0.33, 0.66, 22.34$.  The vectors are the SPGs ($90^o$ rotated) overlaid on the polarized synchrotron intensity map to show how the structure of $|P|$ is changed with respect to the relative scale  $L_{eff}/L$. }
\end{figure*}

We then apply the Faraday Rotation module for $\nu = 0.85, 1.2, 7 GHz$ similar to what we did in \S \ref{sec:limiting}, which corresponds to $L_{eff}/L\sim 0.33, 0.66, 22.34$. From Fig \ref{fig:spg3d-illus}, we clearly see the gradient structure depends on the relative scale $L_{eff}/L$. For example, on the left of Fig \ref{fig:spg3d-illus} only the closest 1/3 of the magnetic field information from the partially rotated cube is effectively projected onto Q and U. Therefore, the gradients from $|P|$ are quite uniformly pointing to the horizontal axis. As the frequency increases to $1.2GHz$ (Fig \ref{fig:spg3d-illus}), the second part of rotated cube, which contains the $30^o$ rotated fields, contributes to the Stokes parameters. As a result, the rotation of  the vectors is observed. However, the weighting of gradients that are further from the observer is generally smaller than of those which are closer to the observer. We, therefore, do not expect the $30^o$ rotated structure to much change the intensity structure. Due to the fact that gradients are linearly added along the line of sight, we can see from the middle of Fig \ref{fig:spg3d-illus} that obvious structures are found oriented $30^o$ away from the horizontal axis. This can be explained by the fact that the resultant SPGs are the linear combinations of sliced gradients along and $30^o$ away from the horizontal axis. In the right panel of Fig \ref{fig:spg3d-illus} we see the case when $L_{eff}/L \gg 1$. This means that the $30^o$ part is {\it fully}  contributing  to the Q,U. As a result, the direction of the gradients are like the average of the horizontal and $30^o$ rotated gradients.

We note that additional information can be obtained by calculating the SIGs of the difference of the total minus  the polarized intensities. This measure tests the magnetic field in the region $[L_{eff}, L]$. This can be complimentary to the SPGs within the studies of 3D magnetic field structure.  We include a preliminarily recipe using in SPDGs in the Appendix \ref{sec:spdg}, but the detailed study is going to be provided elsewhere.

\section{Discussion}
\label{sec:dis}
\subsection{Combining synchrotron gradients and polarization measurement}
\subsubsection{Synchrotron intensity gradients + polarization measurements}
In our earlier paper, i.e. in LYLC17 the synchrotron intensity gradients (SIGs) have been shown to be a  reliable technique for tracing magnetic field (LYLC17).  Unlike tracing magnetic fields with polarization directions, the SIGs are not subject to Faraday rotation and therefore provide do not require multiple frequency measurements to compensate for the Faraday effect. Additional advantage is that measuring intensity is easier than polarization. 

Combining measurements of polarization with the SIGs provides severals ways to obtain synergy of the two measurements. First of all, combining the magnetic field obtained with synchrotron polarization and the SIGs it is possible to increase the reliability of magnetic field tracing. In terms of CMB studies, the information from the SIGs can be used as {\it a prior} for the  establishing the synchrotron polarization directions\footnote{In LY18 we have discussed the possibility of using velocity gradients calculated with HI data for this purpose.}. Combining the SIGs and the polarization is very advantageous, as we do not expect the gradients of CMB intensity to correlate with the directions of synchrotron polarization. Second, the SIGs can help to correct the maps of polarization measured with just one frequency for the Faraday rotation effect. The great advantage of this approach is that no additional measurements are necessary: the same $Q$ and $U$ data will be analyzed in two {\toreferee synergistic} ways. The SIGs require block averaging (YL17a), which decreases the resolution of the map. If the Faraday rotation arising from the nearby galactic regions, the decrease of the resolution is not a problem and the polarization map can be nicely corrected for the Faraday rotation. Moreover, as a result of this correction a map of the Faraday rotations can be obtained. This map can not only inform as about the parallel to line of sight magnetic field strength. To such map one can apply the FRG technique that we described in this paper to get additional information about the magnetic field directions that can be compared to that obtained with the polarization map. The  variations in the directions obtained by the three techniques provide the information about the thermal electron densities that affect the FRGs. 

\subsubsection{Synchrotron polarization gradients + polarization measurements}

All what we have discussed in terms of combining the SIGs and the polarization measurements is applicable to combining the SPGs  with the direct synchrotron polarization measurements for $L_{eff}>L$. Indeed, for this setting one can use the SPGs for (1) improving the accuracy of synchrotron polarization measurements, (2) distinguishing the polarization from synchrotron from the CMB polarization\footnote{The use of SPGs may be advantageous compared to the SIGs as the former are not affected by the gradients from unpolarized radiation.} , (3) provide Faraday rotation maps that can be analyzed to get additional information, e.g. about the parallel  to the line of sight component of magnetic field  and the distribution of thermal plasmas along the line of sight.

The case of $L_{eff}<L$ provides a new direction for studying magnetic fields in 3D. Depending on the correlation properties of magnetic field at the scales sampled by gradients the regions sampled by the SPGs and synchrotron polarized intensities may coincide or may not coincide (see LP16). In the former case point "c." from the section above is applicable but for the probing of magnetic fields and thermal plasmas not through the entire volume, but up to the boundary given by $L_{eff}$. In the latter case, the difference opens ways for getting additional magnetic field information. We shall elaborate these possibilities elsewhere.

\subsection{Studying 3D magnetic field distribution}

\subsubsection{3D structure with synchrotron polarization}

As soon as $L_{eff}$ is less than $L$ the information about the 3D magnetic field {\toreferee distribution} gets available with the SPGs and the SPDGs. Comparing the results obtained with two techniques one can get insight into the distribution of thermal electrons. It also allows to focus studies of magnetic field on particular regions with higher ionization. For instance, if there is a HII region along the line of sight, by comparing the SPGs and the SPDGs measures, one can potentially get insight into the magnetic field structure of the HII region. We intend to try this approach elsewhere. 

The value $L_{eff}$ depends on the Faraday rotation turbulent dispersion or the Faraday rotation depolarization arising from the regular magnetic field. As we get more information about the magnetized turbulent ISM, we get better correspondence between $L_{eff}$ and the line of sight distribution of magnetic field available through the two aforementioned gradient techniques. For instance, the ways of studying Faraday {\toreferee dispersion} using correlations of the synchrotron polarization $P$  and its derivative $dP/d\lambda^2$ ,  are described in LP17.

\subsubsection{Synergy to the Faraday tomography method}

The technique that we proposed in this paper is complementary to the technique of Faraday Tomography proposed by Burn (1966). The latter technique  is  aimed  at  obtaining the information of intrinsic polarization $P_i$ as a function of Faraday depth (see \citealt{BB05}). The method created by Burn (1966) treated Eq (\ref{polar1}) as a Fourier transform of the Faraday dispersion function $F(\phi) = P_i dz/d\Phi$  with the variable $\Phi \propto \int_0^z dz nH_z$, which is the Faraday depth. With conditions for resolution and the effective length-scale condition Eq \ref{eq:el} being satisfied, one can perform the inverse Fourier transform on $\lambda^2$ and to obtain a 3D structure of the Faraday dispersion function.

A significant advantage of our technique (\S \ref{subsec:recipe}) compared to the Faraday tomography is that our technique does not require so many measurements.\footnote{The Nyquist criterion for reconstructing the Faraday dispersion would be in the order of hundreds. Potentially if one consider the sparse properties of $\lambda^2$, the number of measurement may be reduced to $\sim 30$ channels through the compressive sampling theory (Feng Li. et al .2011b). We explore this way of study elsewhere.} 

A serious limitation of the Faraday tomography is related to its treatment of turbulent magnetic fields. In general, the line of sight component of magnetic field in turbulence is expected to change {\kh its} sign. As a result, the Faraday depth $\Phi$ is also changing its sign along the line of sight, {\toreferee i.e. in the case of field reversals the estimation of LOS magnetic field through Faraday depth gets ambiguous.} This presents a fundamental problem\footnote{\toreferee One thing that Faraday tomography relies on is the “LOS distance” traced is in the form of Faraday depth $\Phi\propto \int_0^z dz' \rho(z') B_\| (z')$. If in case the magnetic field is changing signs along the line of sight, the “distance” and the corresponding tomographic slices are in the wrong order  (See Ferriere 2016). This problem has been discussed in Lazarian \& Pogosyan (2016). Comparatively, the SPG relies on the effective distance $L_{eff}$ that does not depend on the sign of B and is constructed by each frequency map providing a reliable LOS distance determination method compared to the Faraday tomography distance through Faraday depth.} for the Faraday tomography. Our gradient approach does not have this sort of limitations and is intended for studying realistic turbulent magnetic fields. The turbulent magnetic field, in fact, provides the real space 3D boundary, which allows the one to study magnetic field up to the boundary $L_{eff}$, the actual extent of which is varied by changing the wavelength. 

We believe that the synergy of the two techniques can be valuable in the future. First of all, we see ways of improving the Faraday tomography. Simple considerations suggest, for instance, that having Position-Position-Frequency rotation cube one can calculate gradients within this cube and those gradients should be perpendicular to the mean magnetic field for the slice of the cube.  We plan to explore these possibilities elsewhere. 

\subsection{Towards a universal gradient technique}
\label{subsec:gradtech}

\subsubsection{General description}
The techniques in this paper that use  polarization gradients should be viewed as a part of the gradient technique that employs modern undrstaning of MHD turbulence.  At first, gradients of velocities are with the Velocity Centroid Gradients (VCGs) (Gonsalvez-Casanova \& Lazarian 2017, Yuen \& Lazarian 2017ab) based on the fact that velocity eddies are elongated along local magnetic field directions (GS95). Later application of the Velocity Gradient Technique include the Velocity Channel Gradients (VChGs) and Reduced Velocity Centroid Gradients (RVCGs) {\kh which} are also explored in LY18

In a separate development, the gradients of the synchrotron emission (SIGs, LYLC) are shown to be capable of tracing magnetic field directions. In the present paper we introduced a few new gradient measures dealing with the synchrotron polarization. The SPGs are mostly sensitive to the polarization fluctuations arising from the variations of synchrotron emissivity, while SPDGs are more focused on the polarization fluctuations arising from the Faraday rotation. For the Faraday rotation screen, the SPDGs transfer to the Faraday Rotation Gradients (FRGs). 

The list above does not exhaust the useful gradient measures. For instance, in Appendix \ref{sec:Zgrad} we discuss the Zeeman Gradients (ZGs) that are very similar in properties to the FRGs. {\kh In earlier discussion} Faraday tomography, we suggested that gradients can be a useful tracer of perpendicular to the line of sight magnetic field direction within this technique. Moreover, applying the block averaging that we developed initially within the VCG approach in YL17a to the intensity gradients we introduced in YL17b, a new technique of Intensity Gradients (IGs)  becomes available for the magnetic field studies.

\subsubsection{Shared foundations and shared procedures and synergy}

All these gradient techniques are based on {\toreferee the same} property of anisotropic MHD turbulence, {\toreferee i.e. turbulent eddies elongated along {\it local} magnetic field }, and use similar procedures for their calculation. The close relation of different gradient techniques is easy to understand. Indeed, magnetic and velocity fluctuations enter the expressions for Alfvenic turbulence in a symmetric way.  This is not generally true about density fluctuations. The studies in Beresnyak et al.  (2005), Kowal et al. (2007) show that for high sonic Mach number $M_s$ the density clumps with a relatively low filling factor dominate both the spectral slope and its anisotropy.  At the same time, turbulent density filling most of the volume exhibit the GS95 scaling with  density passively transferred by the turbulent velocity field.  This opens a way to trace magnetic fields using density gradients, provided that we can identify and remove the volumes of the clumps. This is the essence of our Intensity Gradients (IGs) (see YL17b) technique that has its unique complementary features.  Indeed, the IGs can be used together with velocity gradients in order to identify shocks and regions of gravitational collapse. In some cases when the spectral resolution is not adequate or even not available (as in the case of dust emission), the IGs themselves can be used to get a rough picture of magnetic field distribution. 

We note that the IG technique should be distinguished from the Histograms of Relative Orientation (HRO) technique in Soler et al. (2013) (see also Soler et al. 2017).   In the latter technique, the statistical correlation between the relative orientation of density gradients and magnetic field is sought as a function of column density. Unlike the IGs, the HRO is unable to trace magnetic fields and shocks spatially. HRO and the IGs provide different types of information and are complementary. Indeed, the IGs use the block averaging and are able to trace magnetic field. 

Therefore, advances in one technique entail advances in other techniques. For instance, new procedures of averaging, e.g. moving window procedure and angle constraint procedure, have been tested for the VChGs in LY18. However, these procedures are also applicable of improving the accuracy in other gradient techniques. Similarly, we showed numerically in LYLC that, it is important to filter out the contribution of large scale eddies to trace magnetic fields with SIGs in super-Alfvenic turbulence. We expect the same filtering allows other types of gradients to trace magnetic field in super-Alfvenic turbulent media. 

We also would like to stress that the machinery developed to one gradient technique is usually applicable to other gradient techniques. For instance, the block averaging, originally suggested for the VCGs (YL17a) was later used for other techniques, including synchrotron-based ones. Similarly, more recently discussed techniques of measuring Alfven Mach number $M_A$ by analyzing the distribution function of the gradient directions within a block (Lazarian et al. 2018) can be used to obtain $M_A$ using the synchrotron-based  techniques, i.e. SIGs, SPGs, SPDGs. A significant advantage of the new techniques that we introduced in this paper, namely, the SPGs and the SPDGs, is that the distribution of $M_A$ can be obtained in 3D.

By analyzing the gradient measures from different tracers, one can also study the magnetic field structures and other physical properties from different regimes of the same region. For instance, both total line emission and dust emission can be used for calculating the IGs. Combining different measures, one can get additional information. For example, combining various measures for velocity gradients and the IGs, one can identify shocks in diffuse media and regions of gravitational collapse in molecular clouds (YL17, LY18). 

\subsection{Gradients and the search of cosmological B-modes}

\subsubsection{Polarization-gradient of polarization alignment effect}

Polarization of the galactic foreground is the most serious contaminant for studying B-mode polarization. Our study indicates that there is a very important difference between the properties of the gradients of {\toreferee polarization} related to the 2 types of emission. In particular, the gradients of the foreground polarization are expected to be well aligned with the direction of the foreground polarized emission.\footnote{This statement is getting less accurate if the foreground fluctuation arise from the {\toreferee super-Alfvenic} turbulence. In this situation, the difference of the additive properties of gradients and the Stokes parameters can induce deviations. However, both theoretical arguments and the analysis of observational data testify in favor of {\toreferee sub-Alfvenic} turbulence at high galactic latitudes (see Kandel et al. 2018, Lazarian et al. 2018). For such a turbulence our present study testifies that a very good alignment between the polarization and gradient direction should be present.}  

Very important that this property of gradient alignment with the polarization is a universal property of polarized foregrounds and it is correct not only for the synchrotron foreground contamination. For instance, we discuss in Appendix \ref{dust_pol} that the gradients of polarization from aligned dust are expected to be aligned with the dust polarization directions. 

The effect of the "{\toreferee \it Gradients of Polarization Alignment}" opens a new attractive possibility of separating the foreground polarization from that of the cosmological origin. The presence of the aforementioned effect in the data indicated the presence of the foreground contamination. The structure of the polarization gradients that are directionally correlated with polarization can be used as a prior in separating the polarized contamination of both from synchrotron and dust from the polarized CMB. We will discuss the detailed procedure of the corresponding analysis elsewhere. 

\subsubsection{Synergy with the velocity gradients}

In addition, the analysis of velocity gradients can provide the expected directions of magnetic fields using HI radiation.\footnote{Dust and gas are expected to be well mixed (see Draine 2010). Therefore the magnetic field revealed with the gradients is the same magnetic field that aligns the dust and the intensity of the dust emission is proportional to the 21 cm emission at high latitudes.}  This is similar to the suggestion in Clark et al. (2014, 2015) of using the filamentary structures in velocity channel maps in order to get predictions of the polarization arising from the dust at high galactic latitudes. We believe that the aforementioned filamentary structures are the result velocity crowding and are related to the velocity gradients.  In our collaborative work with S. Clark we are  providing a comparison of the two approaches. We note that Clark (2018) introduced a measure of magnetic tangling that can be obtained through the analysis of the filamentary patten. In Lazarian et al. (2018) we related the dispersion of the gradients with $M_A$ which is a measure of magnetic disorder within MHD turbulence theory.  In fact, in the same paper we related also the dispersion of dust polarization degree to $M_A$. This provides a way of not only predicting the direction of the galactic dust polarization but also predicting the expected degree of dust alignment. 

\subsubsection{Gradients in ISM and star formation studies}

A classical picture of the ISM includes the co-existence of different phases (see Draine 2010). These phases are magnetized with the synchrotron emission arising from hot and warm ISM phases. To understand the connection of magnetic fields in the different phases it is good to use different tracers. Gradients provide such a possibility. For instance, velocity gradients can trace magnetic fields using  various emission and absorption lines.  In YL17b and LY18 a possibility of studying star formation using velocity gradients and the combination of the velocity and intensity gradients have been explored. 

It is very important that gradient techniques provide an extremely promising way of studying the 3D magnetic field structure of the diffuse media. In this paper we have discussed the 3D studies for synchrotron polarization gradients. For velocity gradients the 3D information can also be obtained (LY18). For instance, by employing the Milky Way rotation curve, it is possible determine magnetic field directions as those change along the line of sight,  obtain magnetic fields in high velocity HI clouds, study magnetic fields in molecular clouds that are in galactic disk and occlude each other etc.    The first study of the distribution of magnetic fields in HI disk was performed in  Gonsalvez-Casanova \& Lazarian (2018) with the results successfully tested with the starlight polarization from the stars to which the distances were known.  In addition, the use of multiple chemical tracers provides a promising way of probing 3D magnetic field structure of molecular clouds from the edges (e.g. using CO,CN) to the core (e.g. using NH2). The synchrotron-based technique also provides the 3D information, but by using the wavelength dependence of the polarization decorrelation length. This opens new horizons for the 3D galactic magnetic field studies. 

\subsection{New opportunities for magnetic field studies}

We are currently just scratching the surface of a very rich subject. We expect that further studies of gradients can provide us with a lot of new information about magnetic field structure as well as properties of turbulent interstellar medium. The synergy of different gradient techniques is still something that has not been properly explored. A great advantage of the gradient technique is that the unique information can be obtained even with the existing data sets.

\subsubsection{Studying 3D magnetic field structure of distant objects}

In the previous section (\S \ref{sec:recipe3d}), we see that there is a possibility of obtaining the 3D magnetic field {\kh structures} in the Milky way through the use of SPGs. In fact, it is possible to show that the same 3D mapping is possible for the external galaxies. With the recent advancement of interferometry, we can also measure synchrotron radiation in nearby galaxies. The Faraday rotation induced by the Milky Way interstellar medium acts as a distorting screen for the synchrotron polarization direction. However, it does not change the value of total polarization arising from an external galaxy. As a result, unlike the direction of polarization, polarization gradients are not affected by the Milky Way Faraday rotation and can trace magnetic fields {\kh effectively}

As the observational wavelength changes, the Faraday screening effect that we addressed in \S \ref{subsec:pictorial} becomes also applicable to extragalactic objects and supernovae remnants. As a result, the recipe that we addressed in \S \ref{sec:recipe3d}  is also applicable to these objects. This can bring the studies of synchrotron emitting objects to a new level.

\subsubsection{Synergy with other approaches}

We should stress that while gradients provide by their own information on magnetic field structure, the important synergy between the gradients and polarization exist and should be utilized. We briefly discussed the possible use of the SPGs and SIGs together with polarization. Similarly, the velocity gradients can be used together with dust polarization in order to identify the regions of gravitational collapse within molecular clouds. While our study in LY18 shows that this can be done with velocity gradients alone, the additional information obtained with polarimetry is very helpful. \footnote{Our preliminary analysis shows that such regions of gravitational collapse do not constitute a significant part of the molecular clouds that we analyze. This contradicts to the theories which attribute the non-thermal line broadening observed in molecular gas to the consequence of the gravitational collapse.}

We discussed earlier that the HRO technique (Soler et al. 2013) is different from the IG technique that applies our approaches to density fluctuations. The IGs that trace both magnetic fields and shocks and their joint use with HRO is analyzed in (Yuen \& Lazarian 2018, in preparation).

The gradient technique is also complementary to other techniques of magnetic field studies and  can be potentially be used together for extracting other physical conditions. For instance, the anisotropic behavior from the theoretical prediction of magnetized turbulence theory illuminates the correlation function anisotropy method (Lazarian, Esquivel \& Pogosyan 2002) in studying coarse magnetic field structure. The recipe is then elaborated and thoughtfully tested in subsequent papers (Esquivel \& Lazarian 2004, Heyer et al., Esquivel et al. 2005). We have shown that such correlations correctly trace mean magnetic field, but fail to trace the detailed magnetic field structure that is attainable through the use of gradients (see LY17). In the present paper we trace magnetic fields using the gradients of polarized synchrotron intensities. 

The relation between the polarization gradient amplitudes and the sonic Mach number was studied in Gaensler et al. (2011) and Burkhart, Lazarian \& Gaensler (2012). It was shown there that one can find the sonic Mach number $M_s$ by using these distributions. Our present study suggests that using the depolarization effect one can provide such a study in 3D. Moreover, 
as we discussed earlier, the approach that we successfully tried in Lazarian et al. (2018) by relating the dispersions of velocity gradients to the Alfvenic Mach number $M_A$ is expected to be also applicable to the synchrotron polarization gradients. As a result, apart of tracing 3D distribution of magnetic fields, the synchrotron gradients can provide 3D distribution of $M_s$ and $M_A$, which is very valuable for many describing many astrophysical processes, including the propagation and acceleration of cosmic rays (see Lazarian \& Yan 2014) and star formation (see Burkhart 2018).

\section{Summary}
\label{sec:con}
In the paper above we present a new way of tracing of galactic magnetic field using synchrotron polarization. The theoretical foundations of our new technique are routed both in the modern understanding of the nature of turbulent MHD cascade (GS95, LV99) and in the theory of synchrotron polarization fluctuations (LP16). Our study is focused on exploring the properties of two measures that are available from observations, the SPGs and the SPDGs. We explore how the directions obtained with these measures are related to the directions of the underlying magnetic fields. We consider the effect of Faraday depolarization and study both the cases that this effect arises from the turbulent and the from the mean field. Our main results can be briefly summarized as
\begin{enumerate}
\item Both gradient measures, the SPGs and the SPDGs, are perpendicular to the direction of magnetic field and they allow to trace magnetic fields in both the regime of weak and strong Faraday rotation. The direction of the gradients traces magnetic field and does not require correcting
for the effect of Faraday rotation. 
\item 3D mapping of magnetic fields with the SPGs and the SPDGs can be obtained as the wavelength-dependent Faraday depolarization constraints the depth over which the polarized signal is collected.
\item As the polarized radiation intensity provides the magnetic field strength, SPGs and SPDGs provide not only the direction of the magnetic field in the plane of the sky, but also its amplitude. Moreover, the full 3D vector of the magnetic field can be obtained if the SPDGs technique is employed. 
\item The synchrotron polarization gradients are complimentary to other techniques of magnetic field studies, e.g. our technique that make use of velocity gradients. The approaches developed for the velocity gradient technique, e.g. measurement of Alfven Mach number using the gradient distribution, are applicable to the polarization gradients.  
\item The SPGs and SPDGs open new ways to analyze data for the interstellar medium research in the Milky way and other galaxies, as well as for magnetic field studies in inter cluster medium.
\item The present study indicates additional ways of magnetic field studies, e.g. using the gradients of Faraday rotation measure, dust polarization gradients, Zeeman measure gradients.
\item In terms of CMB polarization studies, in particular, in connection with the search of the ellusive B-modes, our study reveals an important effect of the "polarization-gradient of polarization" alignment" which is expected for both the synchrotron and dust polarization and makes the foreground polarization different from that of the one of cosmological origin. This opens new ways of separating the two types of polarization.
\end{enumerate}

\acknowledgments
{\torefereetwo We thank the anonymous referee for his or her kindly help in improving our paper.} We thank Junda Chen, Victor Lazarian and Bo Yang for their careful proofreading. AL acknowledge the support the NSF grant DMS 1622353 and AST 1715754. 

\appendix
\section{Classification of the Faraday rotation regime in LP16}
\label{sec:LP18}

Here we study numerically, how the turbulence statistics of three regimes described in our paper is related  to the theoretical predictions in LP16. This can be viewed as the numerical testing of the LP16 results that we employ in our paper. In particular, we are most interested in the strong Faraday rotation regime described in LP16.

\subsection{Weak regime}

\begin{figure*}[h]
\centering
\includegraphics[width=0.49\textwidth]{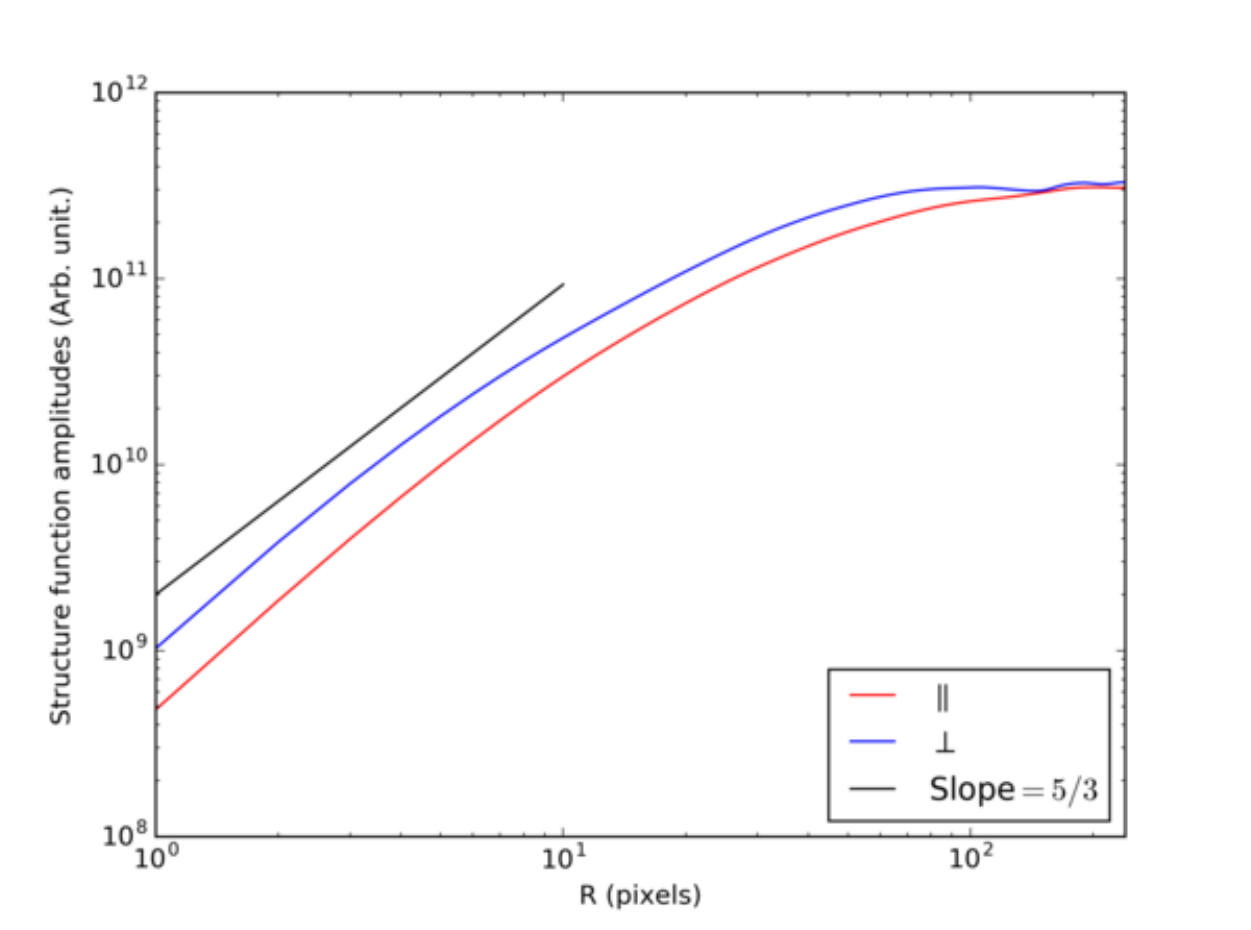}
\includegraphics[width=0.49\textwidth]{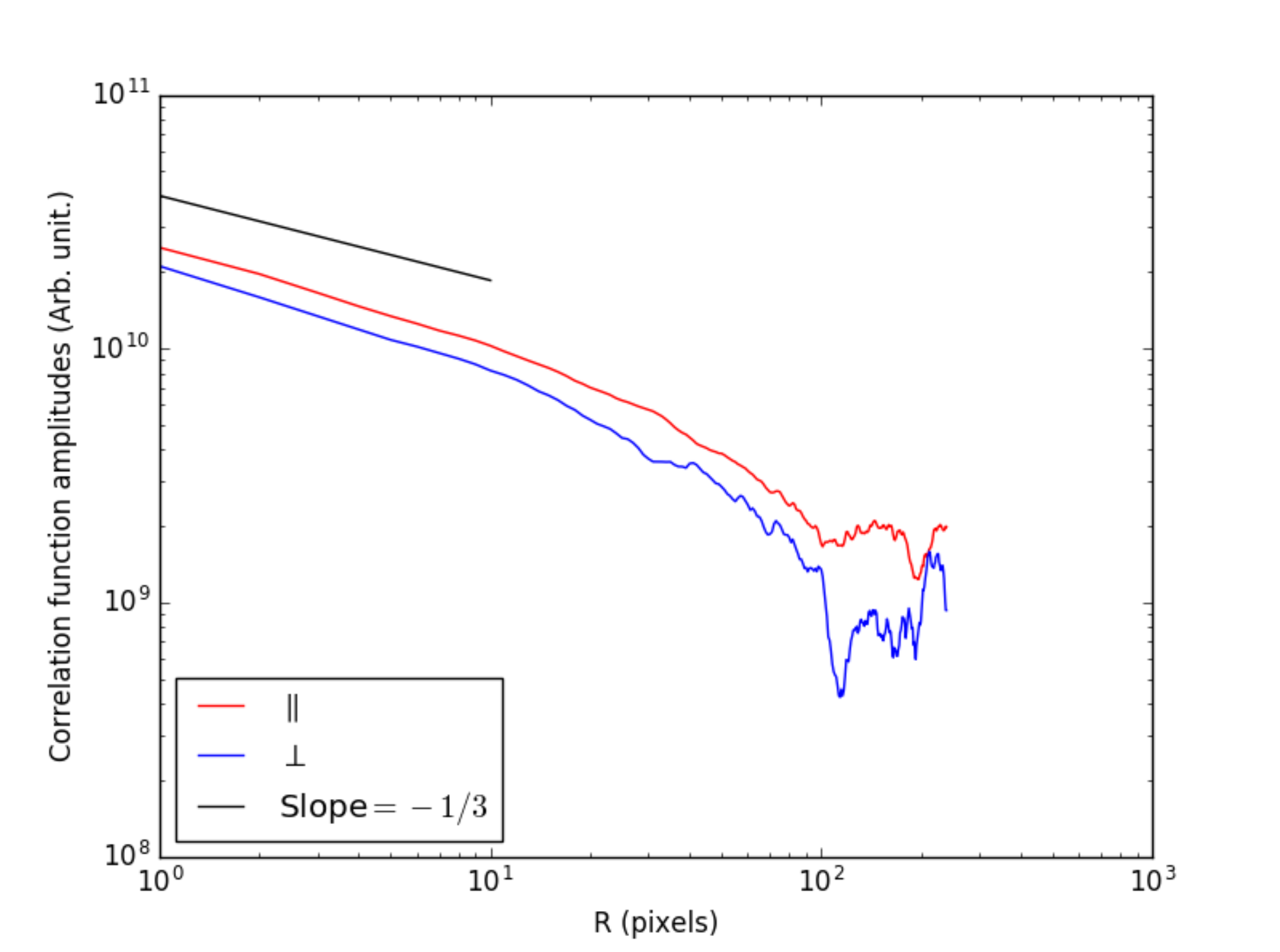}
\caption{\label{fig:LP18-0} {\it Left panel.} The second order parallel and perpendicular structure functions with respect to  2D separation $R$. The weak rotation regime (Left) has an expected dependence of $5/3$ in the structure function. {\it Right panel.} The strong regime exhibits the LP16 predicted  dependence of $-1/3$ in terms of a correlation function. }
\end{figure*}

We study the structure function for both parallel and perpendicular parts of the cube. Fig \ref{fig:LP18-0} shows the second order statistical (structural/correlation) functions with respect to the real space scale. The left of Fig \ref{fig:LP18-0} is the weak case. We see the expected dependence of $5/3$ as we quoted in \S \ref{subsec:spg_weak}.

\subsection{Strong regime}

\begin{figure*}[h]
\centering
\includegraphics[width=0.98\textwidth]{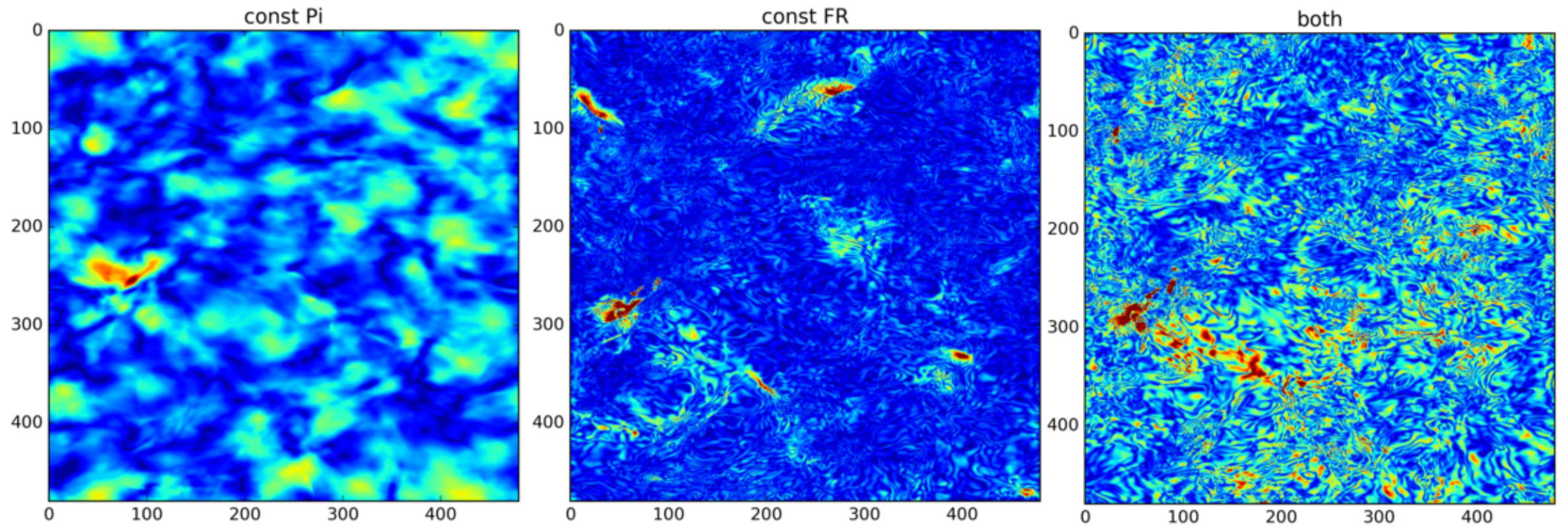}
\caption{\label{fig:LP18-1} The synthetic (obtained with 3D MHD) synchrotron polarization maps {\it Left panel.} with constant $P_i$; {\it Middle panel} constant Faraday rotation term $\Phi$, {\it Right panel.} Both $P_i$ and $\Phi$ vary. Frequency at $\mu=1 GHz$, cube Ms1.6Ma0.53. }
\end{figure*}

The strong Faraday rotation regime is contributed to by both the source term $P_i$ and Faraday rotation term $\Phi$ in Eq (\ref{polar1}). The relative importance of the two terms affect the behavior of both gradients and correlation functions. We therefore, want to test how the relative contributions from the two parts alter the statistics of polarized synchrotron emissions. Fig \ref{fig:LP18-1} shows  how $P_i$ and $\Phi$ affect the polarized maps. To test this, we set one of them to be constant while changing the other. For constant source term case (Left of Fig \ref{fig:LP18-1}), the structure is basically parallel to the mean magnetic field direction (horizontal axis). This is in agreement that the gradients of Faraday Rotation trace magnetic field (see \S \ref{sec:FRG}). For the constant Faraday Rotation case (Middle of Fig \ref{fig:LP18-1}), the structures are more filamentary compared to the constant source term case. This feature is retained in the physical case containing both effects (Right of \ref{fig:LP18-1}), which also exhibits a general alignment towards the mean magnetic field.

The  reduced correlation function (correlation function minus the minimum of it)  for the strong Faraday rotation regime is shown in  Fig \ref{fig:LP18-0}.   The dependence of correlation function on the 2D separation between the points is $R^{1/3}$, which is consistent with the case of intermediate asymptotics obtained in LP16, namely, when $r_\phi> L_{eff}$, the correlation function:
\begin{equation}
\left\langle P({\bf X}_1)P^*({\bf X}_2) \right\rangle \propto r_\phi^{m_\phi/2} R^{-m_\phi/2}
\end{equation}
In our case $m_\phi=2/3$, which means an asymptotic of slope $-1/3$, as shown in Fig \ref{fig:LP18-1}.

\subsection{Numerical artifact from discrete Faraday rotation along the line-of-sight}
\label{subsec:spg_strong}

\begin{figure}[h]
\centering
\includegraphics[width=0.98\textwidth]{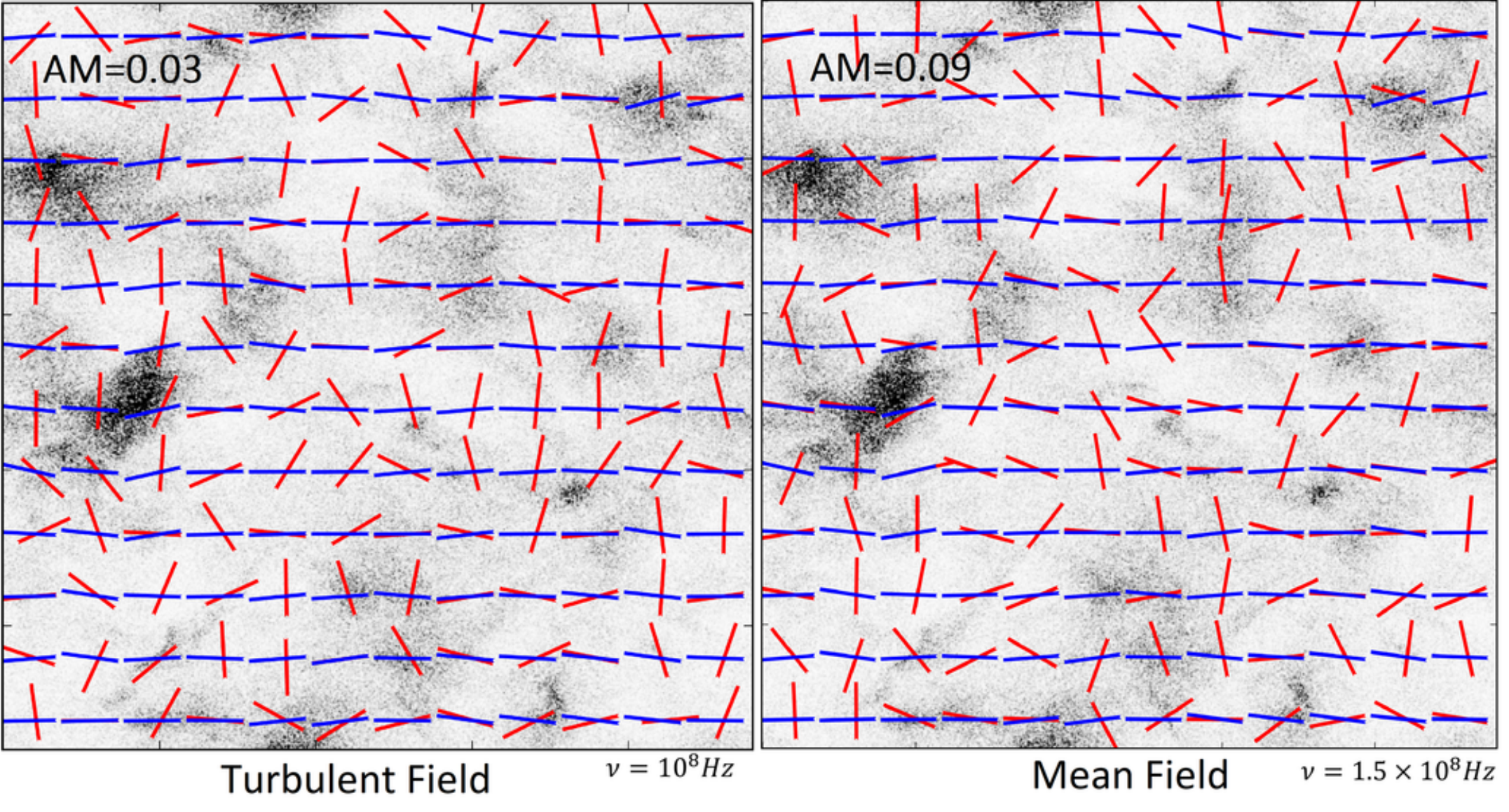}
\caption{\label{fig:LP18-x} In parallel to Fig \ref{f2} and \ref{fig:meanfield} but the polarized synchrotron intensity maps with lower frequencies, which corresponds to the numerical artifact regime (\S \ref{subsec:spg_strong}) in both cases.  }
\end{figure}

\begin{figure}[h]
\centering
\includegraphics[width=0.49\textwidth]{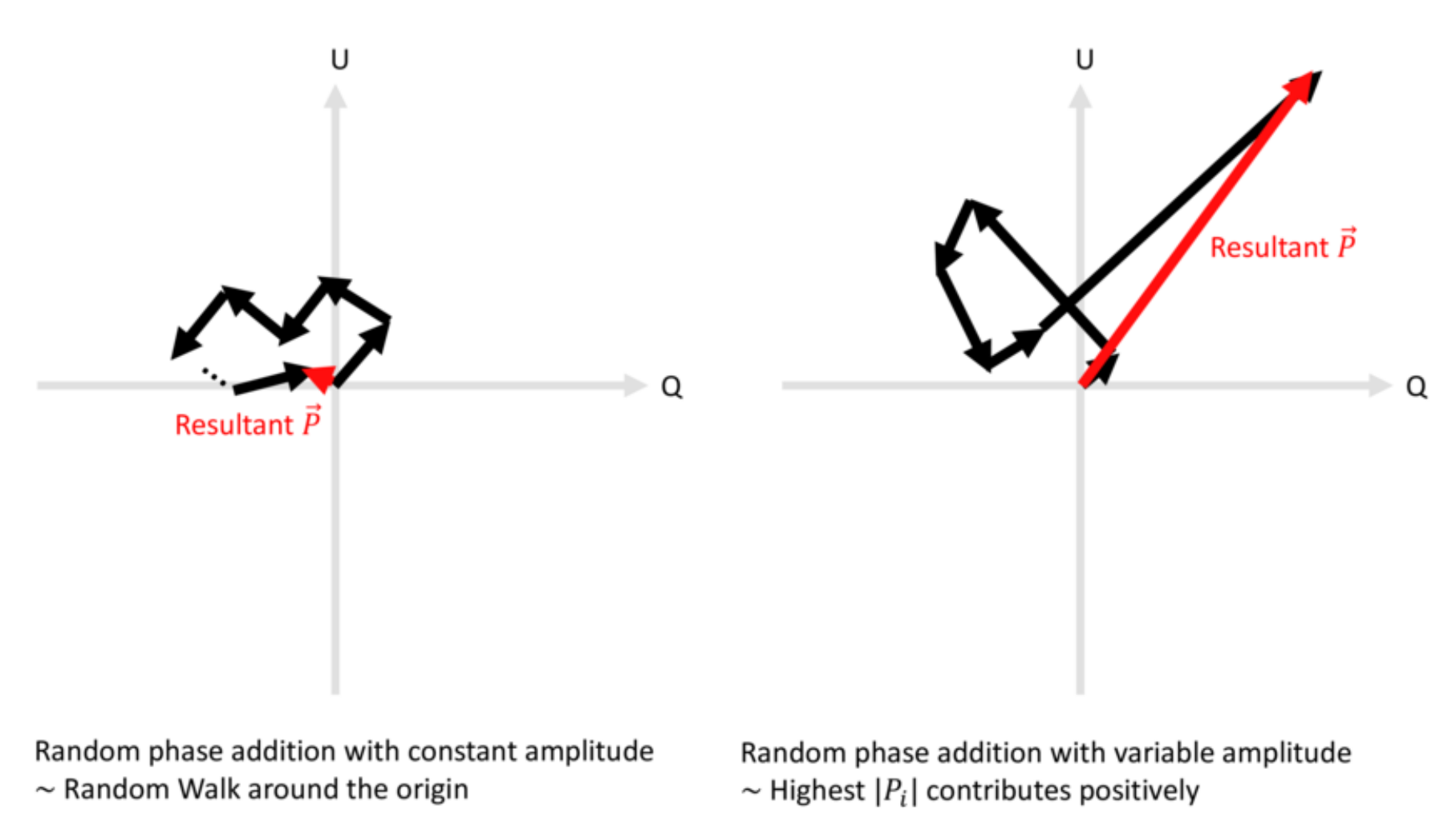}
\caption{\label{fig:LP18-2} The pictorial illustration on how the constant amplitude $P_i=\textit{const}$ and the varying amplitude case (Eq \ref{eq:randadd}) would contribute to $|P|$.  }
\end{figure}

We observe numerical artifacts when the ratio of length scales $L_{eff}/L$ is comparable to the grid size $\Delta z$ of the numerical cube. The striking effect of low frequency synthetic synchrotron emission is that it retains the features of the large scale structure, as shown in Fig. \ref{fig:LP18-x}. In contrast to Fig \ref{fig:amp}, if $\Delta z \ll 1$, the structure of Eq. (\ref{polar1}) looks like the following: Assuming $\theta$ is an uniform random variable in $[0,2\pi]$, Eq. (\ref{polar1}) becomes
\begin{equation}
\label{eq:randadd}
P_{\lambda\gg\lambda_{crit}} \sim \int dz P_i e^{i\theta}
\end{equation}
then the continuous distribution of $P_i$ will guarantee the amplitude of $P$ to drop to zero. However, when there are discrete Faraday source along the line-of-sight, Eq \ref{eq:randadd} becomes
\begin{equation}
\label{eq:randadd2}
P_{\lambda\gg\lambda_{crit}} \sim \int dz P_i e^{i\theta} \delta(\sin(\frac{\pi z}{\Delta z}))
\end{equation}
where $\delta$ is the Dirac delta function. Notice, that the resultant $|P|$ is not zero when $\lambda$ is big. Pictorially we can understand the addition in Eq \ref{eq:randadd2} as illustrated in Fig. \ref{fig:LP18-2}. While the phase is random, the amplitude from the largest vectors is retained, which is exactly what the structure in Eq. (\ref{polar1}) referred to. If the amplitude is set to be constant, the mixing of random phase constant vectors is equivalent to the random walk around the origin. In terms of the polarized intensity amplitude, the varying case reflects how the largest $P_i$ contributes to the map.

\section{Gradients of Zeeman measurements}
\label{sec:Zgrad}
\begin{figure}[h]
\centering
\includegraphics[width=0.49\textwidth]{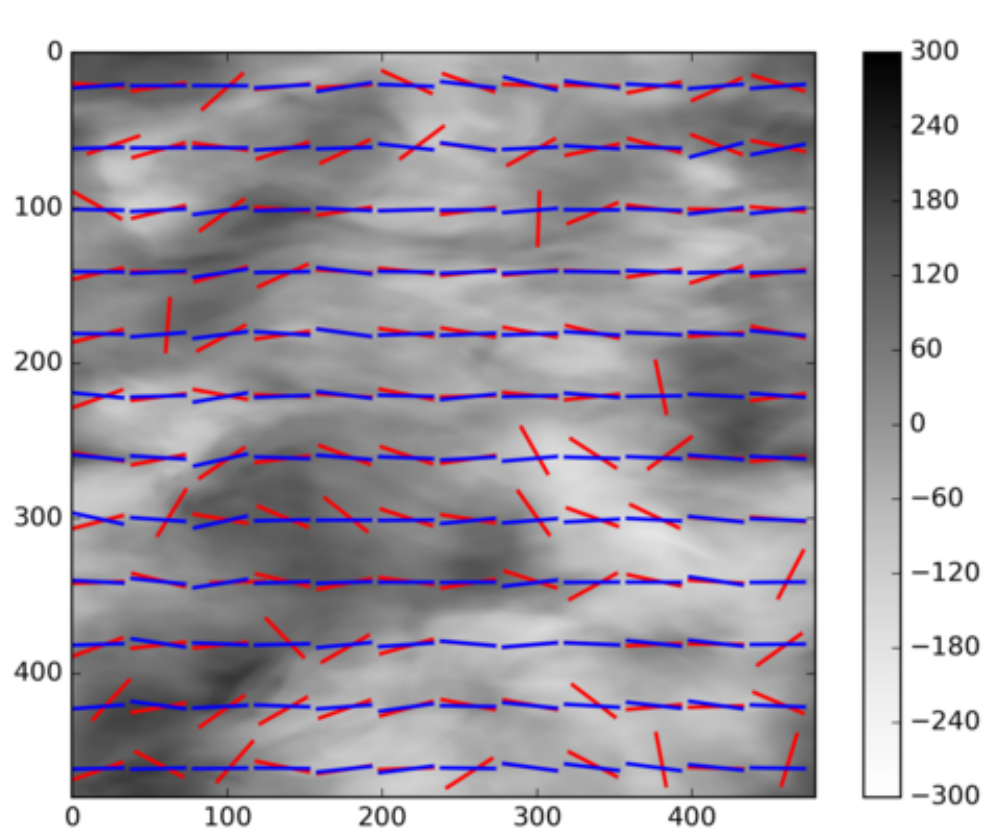}
\caption{\label{fig:Z} The Zeeman gradient (red, $90^o$ rotated) with respect to projected magnetic field (blue) overlaid on Zeeman energy map. }
\end{figure}

The interstellar Zeeman observations provide the direct measurement of the strength parallel to the line of sight component of the magnetic field. The first order energy difference measured due to the Zeeman effect is
\begin{equation}
E_Z^{(1)} = \mu_B g_J B_\parallel m_j
\end{equation}
In special circumstances Zeeman measurement can be available over a patch of the sky. In this situation, the structure of the signal is similar to that of the Faraday rotation. Thus, we expect to have the Zeeman Measure Gradients (ZMGs) directed perpendicular to the plane of the sky magnetic field. In numerical simulation, we mimic the effect of Zeeman effect and assume the quantum quantities ($\mu_B,g_J,m_j$) are constant. Figure \ref{fig:Z} illustrates this effect, the alignment measure is 0.72. 

\section{Gradients of Dust Polarization}
\label{dust_pol}

The observed 2D projection of the emitting volume has the following polarized emissivity (see \citealt{2017arXiv171103161K})
\begin{equation}
\epsilon_{pol}=\epsilon_{Q}+ i\epsilon_U \sim f_{j}(H) (H_x +iH_y)^2
\end{equation}
where $H_x$ , $H_y$ are sky-projected magnetic field components and  for galactic synchrotron $f_{synch}\approx const$ and for galactic dust $f_{dust}=H^{-2}$. Because of the structural similarity of the expressions one should expect that the Dust Polarization Gradients (DPGs), i.e. gradients of $P_{dust}$ , should have properties similar to the SPGs, i.e. gradients of $P$ that we studies in this paper. For the frequencies of dust emission, the Faraday effect is negligible, as a rule. 

The complications arise due to the fact that, unlike relativistic electrons the densities of aligned dust vary significantly. The latter arises both due to dust variations as well as the variations in the dust alignment. The latter effect, however, is not expected to be important for diffuse gas (see \citealt{2007JQSRT.106..225L}). 

The valuable property of dust polarization gradients is that they are expected to be aligned perpendicular to the magnetic field. This property provides a way of distinguish this type of polarization from the polarization of cosmological origin and the polarization from scattered light. Elsewhere we will discuss the procedure of using combining dust emission gradients (or gas density gradients, e.g. obtained via HI emission) and dust polarization gradients in order to separate dust polarization from other sources of polarization. 

\section{Using polarization gradients for {\toreferee studying} 3d magnetic field distribution}
\label{sec:spdg}
\subsection{A brief description of the procedure}
\label{subsec:recipe}

\begin{figure*}[t]
\centering
\includegraphics[width=0.98\textwidth]{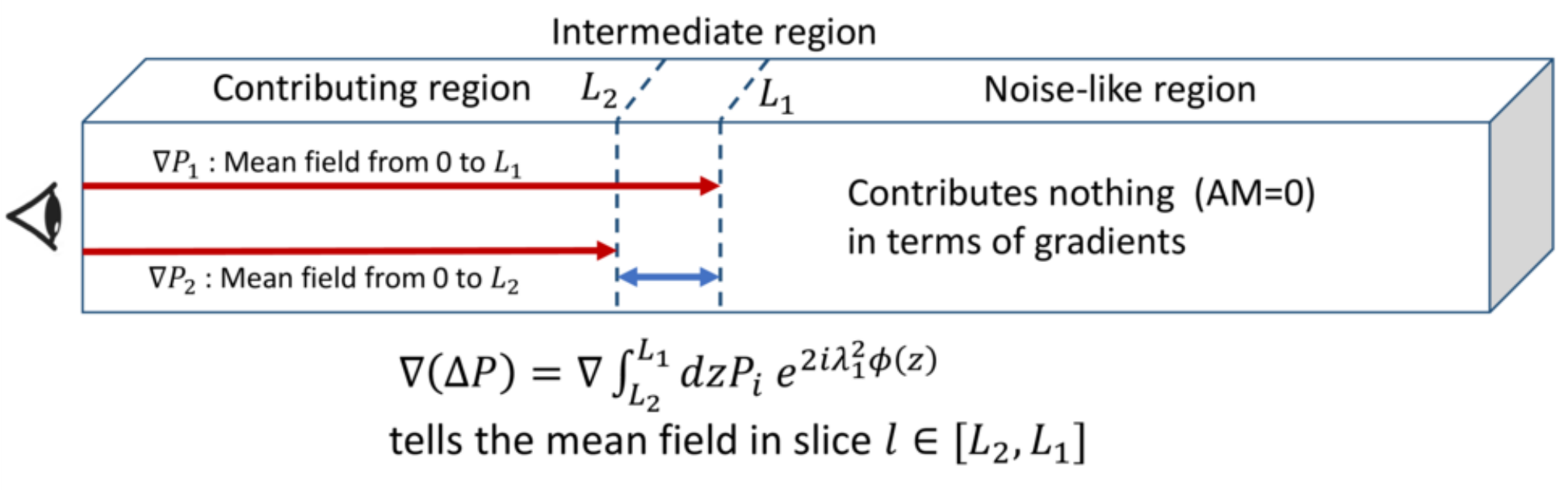}
\caption{\label{fig:3dillus} An illustration of the principle of using SPDGs to construct the 3D magnetic field structure.  The differences of $\nabla|P|$ and two neighboring $\lambda$s allows the {\it mean} magnetic field structure between the slice $[L_2,L_1]$ to be estimated. As long as the slice is thin, the approximation is accurate to the first order. }
\end{figure*}

The discussion of SPDGs in \S \ref{sec:FRG} was focused on using of the FRGs as a synergistic way of studying projected magnetic field direction together with SPGs. From \S \ref{sec:recipe3d}, SPGs represent the cumulative projection of magnetic field structures  for $L_{eff}<L$. Since gradients are linearly added along the line of sight, it is natural to consider the differences of the SPGs from two frequencies $\lambda_{1,2}$, where $\Delta \lambda = \lambda_2-\lambda_1 \ll \lambda_1$. The  difference of SPGs over the change of $\lambda^2$  contains the information of magnetic field projection between $[L_{eff}(\lambda_2),L_{eff}(\lambda_1)]$ (See Eq \ref{eq:el}). In the limit of $\delta \lambda \rightarrow 0$  the difference presents the SPDG measure. We shall denote $L_{i}=L_{eff}(\lambda_i)$ for the remaining part of this section. 

We pictorially illustrate our method in Fig \ref{fig:3dillus}. According to LP16, only a distance smaller than $L_{eff}$  contributes to the polarization measurements taken by the observer, i.e. to  Stokes parameters $Q,U$. The ``1-radian condition'' was introduced in LP16 as the condition for decorrelation and this defines $L_{eff}$
\begin{equation}
\label{eq:oneradcon}
\lambda^2\Phi = \lambda^2 0.81\int_0^{L_{eff}} dz n_e H_z\approx 1.
\end{equation}

The objective in this subsection is to explore how to get the magnetic field morphology, both parallel and perpendicular to line of sight in a slice of $\Delta L \ll L$. The latter  condition is satisfied when $\Delta L / L \sim \Delta (\lambda^2\phi) / (\lambda^2\phi) \ll 1$. The Faraday depolarization effect  allows us to write, for a particular $\lambda$:
\begin{equation}
\begin{aligned}
\label{eq:leffP}
P(\lambda) &= \int_0^\infty dz P_i({\bf X}, z) e^{2i\lambda^2 \Phi(X, z)}\\
&\approx \int_0^{L_{eff}(\lambda)} dz P_i({\bf X}, z) e^{2i\lambda^2 \Phi(X, z)}
\end{aligned}
\end{equation}
The spatial gradients of Eq \ref{eq:leffP} are just SPGs and provide the {\it cumulative} contribution of $P_i$ plus the Faraday rotation factor along the line of sight. 

The Polarization equation for both $\lambda$s (Eq. (\ref{polar1})) are not that different if $\Delta \nu = c/\lambda^2 \Delta \lambda$ is small. Following the discussion in \S \ref{sec:recipe3d}, we split Eq. (\ref{polar1}) into two parts and assume the noise-like part ($z>L_{eff}$) does not contributes to the average of the polarization map. Thus the differences of $P(\lambda_1)$ and $P(\lambda_2)$ is simply:
\begin{equation}
\begin{aligned}
\label{eq:diffP}
\Delta P 
&\approx \int_{L_2}^{L_1} dz P_i({\bf X}, z) e^{2i\lambda^2\Phi(X, z)}
\end{aligned}
\end{equation}
The {\it plane-of-sky direction of magnetic field} between $l\in [L_2,L_1]$ is then given by 
\begin{equation}
\label{eq:pos3d}
\nabla \frac{d|P|}{d\lambda^2} \sim \lambda^{-2} \nabla |P(\lambda_1)-P(\lambda_2)|
\end{equation}

To get the full 3D vector of magnetic field we also need to obtain the {\it inclination angle of magnetic field relative to the LOS direction}, which requires an estimation of magnetic field strength in between $[L_2,L_1]$. From LP16 we know that the quantity $dP/d\lambda^2$ provides the estimate of the Rotation measure, which is linearly proportional to the magnetic field strength:
\begin{equation}
B_z \sim \frac{4\pi}{\Delta L\langle n_e \rangle} \int_{L_2}^{L_1} dz n_e H_z  \sim \frac{4\pi(P_2-P_1)}{(\Delta L\langle n_e \rangle)(\lambda_2-\lambda_1)}
\end{equation}
The strength of the magnetic field component perpendicular to the line of sight $B_{POS}$ can be obtained from the polarized synchrotron intensity, i.e.
$B_{POS}\sim \sqrt{\delta P}/{\Delta L}$, where $\delta P$ is the difference of polarized synchrotron intensity arising from the $[L_2,L_1]$ slice.

The inclination angle of magnetic field relative to the LOS direction can be approximated by:
\begin{equation}
\label{eq:los3d}
\tan \phi = \frac{B_{POS}}{B_z}
\end{equation}
The three dimensional magnetic field can then be constructed when the plane-of-sky gradient orientation $\theta_{grad}$, inclination $\phi$ and total field strength $B_{tot} = (B_{POS}^2+B_z^2)^{1/2}$ are available. By using different  $\lambda_{1,2}$ with small enough $\Delta \lambda$, the 3D  field structure can be obtained. \footnote{The advantage of using the modulus $|P|$ over the cross-intensity $X$ (See \S \ref{sec:limiting}) is evident, since there is an additional phase factor for the complex polarization difference (Eq \ref{eq:diffP}). Considering carefully the rotation of $Q,U$ due to this phase factor, one shall realize the modulus $|P|$ acquires no additional factor while that for cross-intensity $X$ has a factor in the form of $e^{2i\lambda_1^2 \Phi(X, z)}$.}

\begin{figure}[t]
\centering
\includegraphics[width=0.49\textwidth]{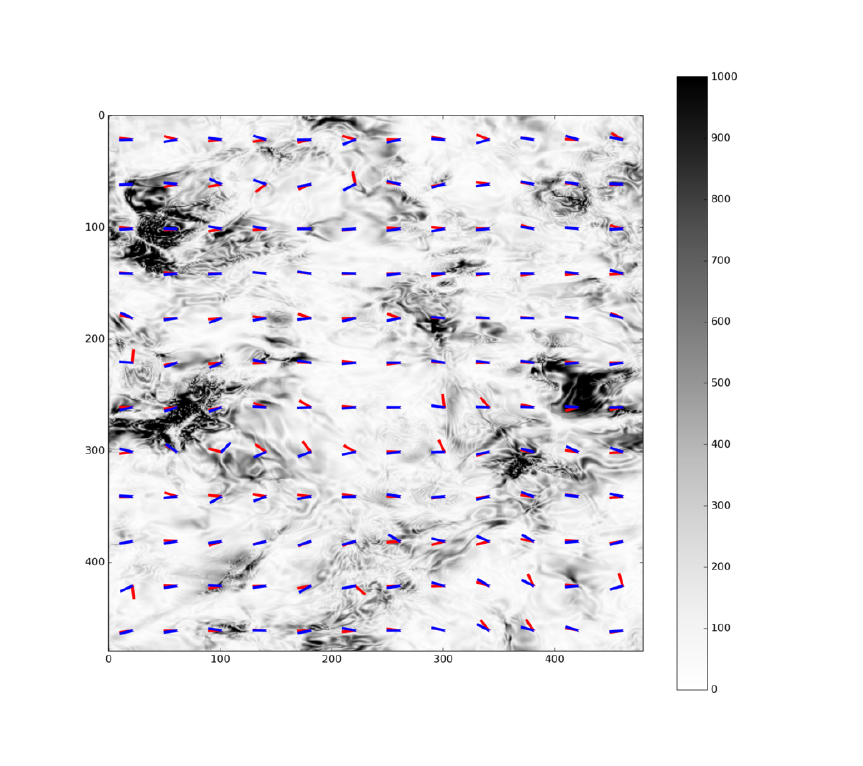}
\caption{\label{fig:sliceB} The sliced SPDGs (rotated $90^o$, red) compared with the sliced projected magnetic field (Blue) overlaid on the $d|P|/d\lambda^2$ when $\nu=1Ghz$. The alignment measure is $0.73$ }
\end{figure}
\begin{figure}[t]
\centering
\includegraphics[width=0.23\textwidth]{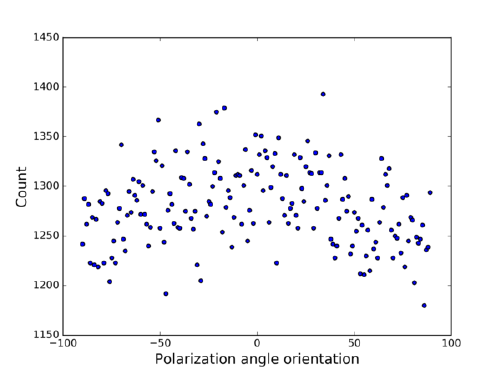}
\includegraphics[width=0.23\textwidth]{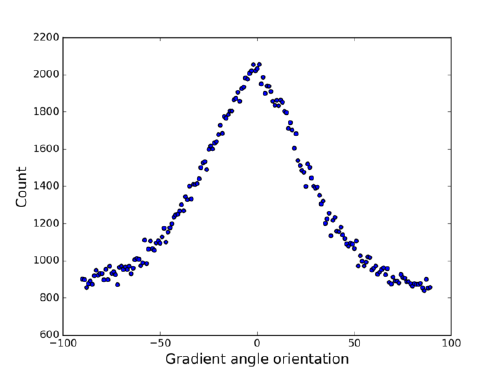}
\caption{\label{fig:orient} The angle orientation histograms for polarization angle computed through difference of Stokes parameters (Left) and the SPDGs orientation (Right). The zero degree corresponds to the horizontal axis. }
\end{figure}


\subsection{Tracing the distribution of plane of sky magnetic fields in a slice }
We test our recipe (\S \ref{subsec:recipe}) in the random field case, which corresponds to the calculation we did in \S \ref{sec:limiting}. We pick $\nu=1GHz$ with $\Delta \nu=10Mhz$, which corresponds to the case of $\Delta \Lambda = 0.003$m, $L_{eff} = 4.9$ and $\Delta L = 0.21$ (10 slices over 480 spatial resolution for Ms1.6Ma0.53). The condition $\Delta L/L\ll 1$ is satisfied and therefore the polarization gradients can be computed readily following the treatment in \S \ref{sec:FRG} and \S \ref{subsec:recipe}. The resultant $d|P|/d\lambda^2$ map and its gradients (SPDGs, in red) are shown in Fig \ref{fig:sliceB}. To compare with polarization, we compute the projected magnetic field from the 3D magnetic field data with line-of-sight deepness of $[L_{eff}-\Delta L,L_{eff}]$. Numerically we compute the effective Stokes parameters $(Q_{\Delta L},U_{\Delta L})$ by
\begin{equation}
\begin{aligned}
Q_{\Delta L} &\sim \int_{L_{eff}-\Delta L}^{L_{eff}}  dz Q({\bf X},z)\\
U_{\Delta L} &\sim \int_{L_{eff}-\Delta L}^{L_{eff}}  dz U({\bf X},z)\\
\end{aligned}
\end{equation}
The {\it effective polarization angle $\theta_{\Delta L}$}, which reflects the average magnetic field orientations in the deepness of $[L_{eff}-\Delta L,L_{eff}]$. The alignment measure between SPDGs and the effective polarizations in Fig \ref{fig:sliceB} is $AM=0.73$.

We have shown that we can trace magnetic field using the polarization gradients as $L_{eff}<L$. It is easy to see that one cannot use the direction of of synchrotron polarization to do the same job. For instance, Fig \ref{fig:orient} shows the orientation histogram (See YL17a for a formal definition) of both polarization angles computed through differences of Stokes parameter (Left) and the SPDGs orientation (Right). One can see there is a preferred direction for gradient angle orientation inside a block but not for the polarization. Therefore, we conclude that the gradients provide a unique way of evaluating magnetic field structure that has significant advantages compared to the traditional magnetic field tracing using synchrotron polarization directions. 

\subsection{3D construction}

As we discussed above, we can use data corresponding to different $(L_{eff},\Delta L_{eff})$ to reconstruct the 3D magnetic field  structure. To reduce the complexity, we would use the mean-field results from a 15-slice irregular $\Delta \nu$ PPF cube ranging from $1.011Hz$ to $1.153GHz$. The sample PPF cube would then provide 14 slices of magnetic field with length scale from $L=4.56$ to $L=4.66$ with equal $\Delta L$ spacing of $0.1$. We listed our choice of frequencies in Table \ref{tab:ftable} for reader's reference. The spacing corresponds to about 5 slices out of 480 slices in our cube with total LOS depth of 10pc ($5/480*10\sim 0.104pc$). Therefore, we selected the appropriate slices from the original cube with Stokes-parameter addition {\it within the slice} to obtain the 5-slice POS component and taking 5-slice mean for the LOS component. Under this construction, we can compute the 3D magnetic field by gradients following the recipe in \S \ref{subsec:recipe} and compare it with the mean 3D orientation from the 5-slice. The differences between the mean inclination is about $2^o$.

\begin{table}[t]
 \centering
 \label{tab:ftable}
 \begin{tabular}{c c c c}
$\nu$ (GHz) & $\Delta \nu$ (MHz) & $L_{eff}$ (pc) \\ \hline \hline
1.011 & 11 & 4.56\\
1.022 & 11 & 4.66\\
1.032 & 11 & 4.76\\
1.042 & 11 & 4.86\\
1.053 & 10 & 4.96\\
1.064 & 10 & 5.06\\
1.074 & 10 & 5.16\\
1.084 & 10 & 5.26\\
1.094 & 10 & 5.36\\
1.104 & 10 & 5.46\\
1.114 & 10 & 5.56\\
1.124 & 10 & 5.66\\
1.134 & 10 & 5.76\\
1.143 & 10 & 5.86\\
1.153 & 10 & 5.96\\
\hline \hline
\end{tabular}
 \caption {The irregular $\Delta \nu$ PPF construction.  }
\end{table}

\begin{figure}[t]
\centering
\includegraphics[width=0.49\textwidth]{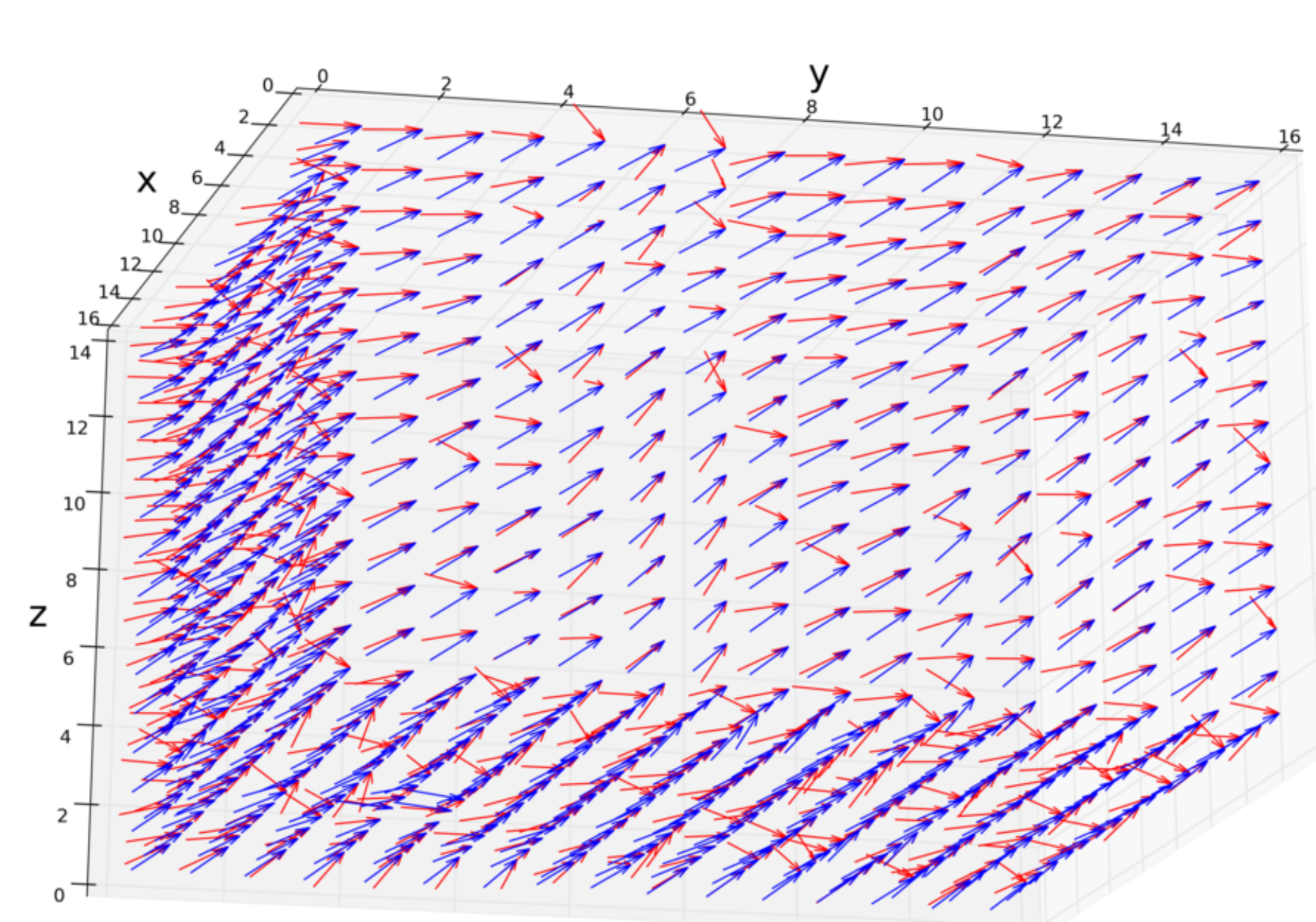}
\caption{\label{fig:b3d} The three-dimensional magnetic field morphology constructed from the real slices spaced $\Delta L=0.1$ (blue) and from our recipe (red). The vectors are clearly aligned to each other.  }
\end{figure}

\begin{figure}[t]
\centering
\includegraphics[width=0.49\textwidth]{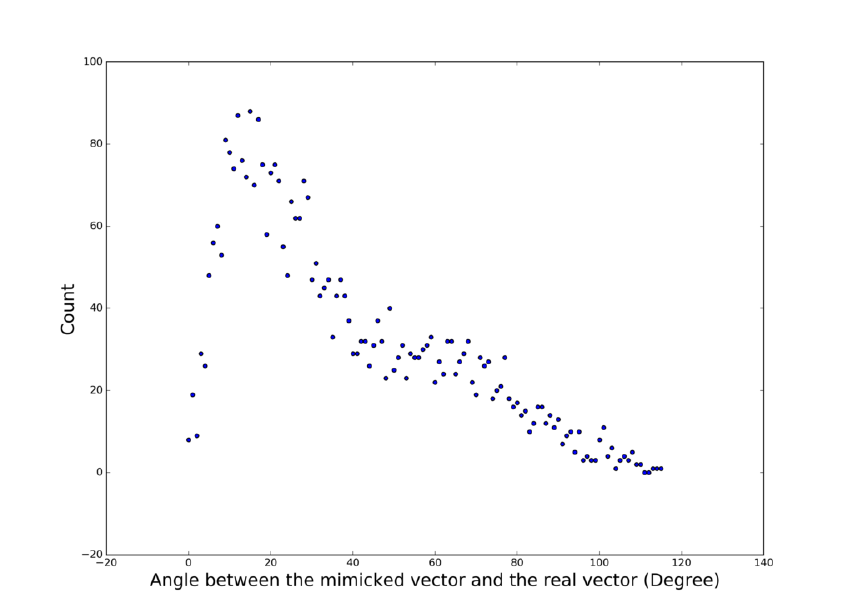}
\caption{\label{fig:b3ds} The 3D relative angle scatter plot by taking dot product on the unit vectors of B-field predicted by the recipe (\S \ref{subsec:recipe}) and the real field.  }
\end{figure}

Fig. \ref{fig:b3d} shows the 3D magnetic field constructed by \S \ref{subsec:recipe} compared to the real 3D structure from the same cube Ms1.6Ma0.53. For the sake of the comparison, we show only the first layer of the vectors for readers to compare. The 3D morphology predicted by the recipe is close to what is in the original data. We also compare the two 3D vector field by their dot product, which gives out the cosines of the relative angle between them. Fig. \ref{fig:b3ds} shows the plot of the relative angle. The peak is at around $15^o$, which means the recipe from \S \ref{subsec:recipe} is  promising as the first step of tracing 3D magnetic field. Further improvements of the technique will be discussed elsewhere. 
 
\end{document}